\documentclass[aps,pra,reprint,superscriptaddress]{revtex4-1}

\usepackage{graphicx}
\usepackage{dcolumn}
\usepackage{bm}
\usepackage{epstopdf}
\usepackage{amsmath}
\usepackage{amssymb}
\usepackage{latexsym}
\usepackage{revsymb}

\begin{document} 


\title{Heating and cooling of electrons in an ultracold neutral plasma using Rydberg atoms}


\author{E. V. Crockett}
\altaffiliation{Present address: Lincoln Financial Group, Boston, MA 02110}
\affiliation{Department of Physics and Astronomy, Colby College, Waterville, Maine 04901, USA}

\author{R. C. Newell}
\altaffiliation{Present address: Applied Physics Laboratory, University of Washington, Seattle, WA 98105}
\affiliation{Department of Physics and Astronomy, Colby College, Waterville, Maine 04901, USA}

\author{F. Robicheaux} 
\affiliation{Department of Physics and Astronomy, Purdue University, West Lafayette, Indiana 47907, USA}

\author{D. A. Tate}
\email[]{duncan.tate@colby.edu}
\affiliation{Department of Physics and Astronomy, Colby College, Waterville, Maine 04901, USA}




\date{\today}

\begin{abstract}
We have experimentally demonstrated both heating and cooling of electrons in an ultracold neutral plasma (UNP) by embedding Rydberg atoms into the plasma soon after its creation. We have determined the relationship between the initial electron temperature, $T_{e,i}$, and the binding energy of the added Rydberg atoms, $E_b$, at the crossover between heating and cooling behaviors (that is, the binding energy of the atoms, which, when they are added to the plasma, neither accelerate or slow down the plasma expansion). Specifically, this condition is $|E_b| \approx 2.7 \times k_B T_{e,i}$ when the diagnostic used is the effect of the Rydberg atoms on the plasma asymptotic expansion velocity. Additionally, we have obtained experimental estimates for the amount of heating or cooling which occurs when the Rydberg binding energy does not satisfy the crossover condition. The experimental results for the crossover condition, and the degree of heating or cooling away from the crossover, are in agreement with predictions obtained from numerical modeling of the interactions between Rydberg atoms and the plasma. We have also developed a simple intuitive picture of how the Rydberg atoms affect the plasma which supports the concept of a ``bottleneck'' in the Rydberg state distribution of atoms in equilibrium with a co-existing plasma. 
\end{abstract}


\maketitle


\section{{INTRODUCTION}\label{intro}}

Ultracold neutral plasmas (UNPs), first observed at NIST in 1999 \cite{kill99}, are systems in which basic atomic processes dominate the plasma evolution process (for a comprehensive picture of recent research on UNPs, see Refs. \cite{kill07} and \cite{lyon17}). UNPs are created by photoionizing cold atoms in a magneto-optical trap (MOT), though they have also been made using translationally cold atoms and molecules in a supersonic beam \cite{morr08,schu16b}. Additionally, cold, dense Rydberg atom ensembles evolve spontaneously into UNPs \cite{rob00,li04,fore18}. 

A question that has received significant attention with regard to UNPs is whether they can reach the strongly coupled regime for either the ions or the electrons. The traditional measure of the strength of the coupling is the ratio of the mean electrostatic interaction energy of a species to its mean kinetic energy, which is parameterized by the Coulomb coupling parameter,
\begin{equation}
\Gamma_\alpha = \frac {e^2}{4 \pi \epsilon_0 a_{\alpha} k_B T_\alpha} \label{gamma}
\end{equation}
($a_{\alpha}$ is the Wigner-Seitz radius for species $\alpha$, which may be electrons, $e$, or ions). Plasmas with $\Gamma_\alpha \gtrsim 1$ are considered to be strongly coupled for species $\alpha$, and manifest long-range correlation behaviors \cite{poh04a}.

For UNPs made by photoionization of atoms in a MOT, the initial ion temperature, $T_{ion,i}$, is the same as that of the MOT atoms and lies in the range 0.1 - 10 mK for plasmas made from alkalis, alkaline earths, and noble gases. The initial electron temperature, $T_{e,i}$, is determined by the laser excess photon energy, $\Delta E$, as
\begin{equation}
\Delta E = \frac{3}{2} k_B T_{e,i} = h \nu - E_{IP}, \label{Teinit}
\end{equation}
where $E_{IP}$ is the atom's ionization energy. Using a narrow bandwidth laser to ionize the atoms, it is possible to make $T_{e,i}$ very small. For instance, in Ref. \cite{chen17}, a value of $T_{e,i} = 0.1$ K was achieved. Additionally, ion and electron densities of up to $\sim10^{10}$ cm$^{-3}$ \cite{gupt07} are attainable in a MOT, and may be as high as $\sim10^{13}$ cm$^{-3}$ in a supersonic beam \cite{schu16b}. 

Based on these temperatures and densities, it would seem relatively straightforward to reach the strongly coupled regime for both electrons and ions. However, the low initial electron temperatures in such plasmas result in high rates for three body recombination (TBR), which scales with ion and electron densities, $\rho_{ion}$ and $\rho_e$ respectively, and electron temperature $T_e$ as $\rho_{ion} \, \rho^2_e \, T^{-9/2}_e$ \cite{mans69}. TBR heats the plasma electrons and results in the formation of Rydberg atoms. In contrast, the inverse process, namely collisional ionization of the Rydberg atoms, cools the plasma electrons, but electron collisions with the atoms can also de-excite the atoms and heat the electrons. Additionally, several other mechanisms heat electrons and ions. For instance, both electrons and ions are subject to disorder induced heating (DIH), and this process typically heats the ions up to $\sim 1$ K in the first few microseconds of the plasma evolution process at higher densities \cite{chen04,lyon17}. 

In this paper, we distinguish the initial temperature values which are set by the experimental conditions from the effective initial electron and ion temperatures, $T_{e,0}$ and $T_{ion,0}$, respectively. It is the thermal energy represented by these latter two quantities that drives the plasma expansion to have an asymptotic expansion velocity 
\begin{equation}
v_0 = \sqrt{\frac{k_B (T_{e,0} + T_{ion,0})}{m_{ion}}}, \label{vZero}
\end{equation} 
where $m_{ion}$ is the ion mass. In an ideal collisionless plasma, $T_{e,0} = T_{e,i}$ and $T_{ion,0} = T_{ion,i}$; however, the heating mechanisms mentioned above cause $T_{e,0}$ and $T_{ion,0}$ to be larger than the initial values set in the experiment. The most straightforward techniques for measuring $T_{e,0}$ and $T_{ion,0}$, including the method we use in this paper to obtain $T_{e,0}$, directly or indirectly extract its value from the plasma expansion velocity using Eq. \ref{vZero} \cite{kul00,sim04,morr08}. 

In general, heating of ions is less significant than for electrons. Consequently, values for $\Gamma_{ion} \gtrsim1$ have been achieved, and there are mechanisms, either proposed or already demonstrated, that may be used to cool the ions in a UNP, and increase $\Gamma_{ion}$. In UNPs made from Sr atoms, laser cooling using the Sr$^+$ resonance line has been reported in Ref. \cite{gorm18}, and it has been proposed that high $\Gamma_{ion}$ values may also be achievable in UNPs which evolve from dipole-blockaded cold Rydberg samples \cite{bann13}. On the other hand, for small $\Delta E$, TBR heats the electrons and results in minimum $T_{e,0}$ values in the range 30 - 50 K, and, at high density, DIH will also cause electron heating \cite{kuz02b}. (Electron correlation heating in UNPs created with initial electron temperature and density such that $\Gamma_e \sim 1$ can also be considered using the threshold lowering model \cite{hahn02}). Because of these electron heating mechanisms, it has been found that $\Gamma_e \lesssim 0.2$ in UNPs \cite{robx02}. 

In contrast with the ability to cool ions in a UNP, there are limited avenues for achieving strong electron coupling in UNPs created from cold atoms in a MOT, due to the strong dependence of TBR rates on electron density and temperature. The NIST group attained $\Gamma_e = 0.13$ \cite{flet07} for Xe plasmas, while the group at Rice University found $\Gamma_e \lesssim 0.2$ for Sr plasmas made by photionization \cite{gupt07}. The group at Colorado State University obtained $\Gamma_e = 0.35 \pm 0.08$ for Rb plasmas made at low density so that TBR heating was minimal \cite{chen17}. On the other hand, UNPs which evolve from Rydberg states of cold NO molecules in a supersonic beam (for a review of these experiments, see Ref. \cite{schu16b}) have been reported to have $T_e \approx 7$ K at a density such that $a_{e} = 360$ nm, implying $\Gamma_e \approx 7$ \cite{morr09}. At present, there is no obvious way to bridge the factor of 1000 difference in density between the MOT experiments and the beam experiments in order to understand the difference in the degree of electron coupling that can be achieved. Moreover, UNPs made by exciting molecular Rydberg states have a number of degrees of freedom that are unavailable to atomic UNPs, further complicating attempts to compare the dynamics of cold atomic and molecular plasmas.

In Ref. \cite{van05}, it was proposed that the electrons in a UNP could be cooled by adding (``embedding'') Rydberg atoms into a UNP. While the experiments reported in this paper were not sensitive enough to detect a change in the electron temperature due to the addition of Rydberg atoms, a related numerical study \cite{poh06a} showed that the experimental results of Ref. \cite{van05} were consistent with modest heating or cooling of plasma electrons by the introduction of Rydberg atoms. The basic idea of the cooling mechanism is that electron-Rydberg collisions either ionize the atom, or leave it in a higher-lying energy state. It has been known for a long time that there exists a bottleneck in the Rydberg state distribution for atoms which co-exist with a plasma \cite{mans69}. At the bottleneck, the atoms have an energy that is lower than the ionization limit by an amount $E_{bn} \approx 4 k_B T_e$, where $T_e$ is the plasma electron temperature. Atoms with this energy are as likely to ultimately ionize due to electron collisions as they are to be de-excited and eventually decay radiatively so that they no longer participate in the plasma evolution dynamics. Therefore, adding atoms which are bound by more than $E_{bn}$ should heat the plasma electrons since the most likely collisions are those that result in the atom being more deeply bound than less, while adding atoms with bound by less than $E_{bn}$ should result in electron cooling.

Here, we present an experimental and numerical study of adding Rydberg atoms to a cold plasma. Specifically, using both direct and indirect measurements of the plasma asymptotic expansion velocity, $v_0$, obtained from the electron and ion time of flight (TOF) spectra, we have observed both heating and cooling of UNPs due to the presence of Rydberg atoms. Additionally, we have measured the critical Rydberg atom energy, $E_b$, that leaves the plasma expansion velocity unchanged for a UNP with initial electron temperature $T_{e,i}$. (In this paper, we will call the value of $T_{e,i}$ for that $|E_b|$ value the crossover temperature for the electrons, $T_{CO}$.) Further, using the ion TOF spectra, we have quantified the amount of heating and cooling that results when the Rydberg binding energy is not at the crossover condition for the plasma electrons and how this depends on the Rydberg binding energy and density. Our experimental results agree with numerical simulations using the Monte Carlo approach. 

The principal results of our study are as follows. First, the crossover temperature of the electrons is related to the binding energy of the Rydberg atoms by the equation $|E_b| \approx 2.7 \times k_B \, T_{CO}$, which corresponds to the condition where the average energy gained by a Rydberg atom in a collision by an electron when the atom is excited is equal to the mean energy lost by the atom when the atom is de-excited. This is different than the bottleneck condition described above \cite{mans69,vriens80,steve75,kuz02b,poh08,bann11}, for which $|E_b| \approx 4 \times k_B \, T_{e}$. Second, when the Rydberg atom binding energy and the electron temperature are not in the crossover condition, the average amount by which the energy of the plasma electrons is increased by a single electron-atom collision is $0.353 \, |E_b| - k_B \, T_e \propto T_{CO} - T_e$. Consequently, the net amount by which the plasma is heated or cooled is proportional to $T_{CO} - T_e$, and also to the number of Rydberg atoms, and the mean number of electron collisions each Rydberg atom experiences during the plasma evolution. Unfortunately, our results suggest that embedding Rydberg atoms into a UNP may have a limited ability to push a plasma into the strongly coupled regime for electrons.

\section{{EXPERIMENT}\label{exp}}

The apparatus used in this experiment has been described previously \cite{bran10,fore18}. We use a rubidium vapor-cell MOT which is capable of trapping up to $1.2 \times 10^8$ $^{85}$Rb atoms at a peak density of $5 \times 10^{10}$ cm$^{-3}$ and a temperature of $\approx 100$ $\mu$K. The atoms are trapped midway between a pair of flat, high transparency copper meshes (the ``field meshes''), which allow the application of small dc electric fields and/or high voltage pulses for selective field ionization (SFI) of Rydberg atoms. The MOT is run continuously, and neither the cooling nor the repump lasers are switched off during the experiment.
The number of atoms in the MOT is obtained from a measurement of the total 780 nm fluorescence power emitted by the atoms, and the FWHM of the spatial density distribution is found by imaging the fluorescence onto a linear diode array. The $e^{-1/2}$ radius of the density distribution is in the range $\sigma_0 = 500 - 600 \ \mu$m. 

The plasma is created (at $t=0$) by a pulse of $\approx$480 nm light from a home-built Littman type dye laser \cite{litt78} (Coumarin 480 dye) which photoionizes a fraction of the $5p_{3/2}$ Rb atoms. The dye laser is pumped by third harmonic light from a 20 Hz repetition-rate Nd:YAG laser (Continuum Surelite), and has a pulse duration of $< 5$ ns. Calibration of the dye laser frequency is achieved using a 0.5-meter spectrometer referenced to the Balmer-$\beta$ line in hydrogen. This enabled us to create plasmas with initial electron temperature $T_{e,i}$ values (as defined in Eq. \ref{Teinit}) in the range 20 to 140 K with an uncertainty $\lesssim 5$ K. The line width of the Littman laser is $\approx 3$ GHz, equivalent to a temperature uncertainty of $\approx 0.1$ K. The Surelite Nd:YAG pump laser is not injection seeded, and while this likely has some impact on the Littman laser line width, it does not affect the precision of the $T_{e,i}$ values as this is limited by the frequency calibration method. (The $T_{e,i}$ uncertainty, $\pm 5$ K, is equivalent to a frequency uncertainty of $\pm 100$ GHz.)

Rydberg atoms are embedded in the plasma at a time $\Delta$ after the creation of the plasma, where $\Delta =$ 25 ns in the experiments described here. The laser used to excite the $5p_{3/2}$ $\rightarrow$ $nd$ transition (where $n =$ 24 - 60) is a narrow bandwidth pulsed laser (NBPL). Light at $\approx 960$ nm from a continuous wave external cavity diode laser (ECDL) is pulse amplified in three dye cells (LDS 925 dye), and then frequency doubled using a KNbO$_3$ crystal \cite{bran10}. The dye cells are pumped by 532 nm light from a second Nd:YAG laser (Continuum YG-661) in which the $Q$-switch is electronically triggered at a precisely controllable delay after the $Q$-switch of the Surelite Nd:YAG which is used to pump the Littman laser. The YG-661 pump laser, like the Surelite, is not injection seeded, and so the NBPL pulses are not transform limited. The fact that the YG-661 is unseeded results in longitudinal mode beating in the 532 nm light which pumps the LDS 925 dye, and this has been found to broaden the wings of the output spectrum of dye amplifiers (see, for instance, Ref. \cite{tric07}). Nevertheless, we achieve a FWHM line width of $\approx 200$ MHz, which we have verified by frequency scanning the 960 nm ECDL and obtaining a spectrum of several states in the $n = 90-91$ Rb Rydberg manifold. This signal was acquired using SFI, and the frequency scale at 960 nm was calibrated using a 1.5 GHz confocal etalon, and at 480 nm using the known spacings of the states which appeared in the SFI spectrum. The NBPL line width is sufficient to resolve well the $j=5/2$ and $j=3/2$ $32d$ states which are separated by 364 MHz \cite{harv77}. (For $n \le 32$ we excite the $j=5/2$ state, but above $n = 32$ we excite a mixture of the two $nd_j$ levels.) 
More importantly, the the shot-to-shot variations in the NBPL spectrum result in only small variations in the number of Rydberg atoms created. The ECDL laser wavelength is measured using a Burleigh WA-1500 wave meter, resulting in a negligible uncertainty in knowledge of the binding energy of the Rydberg atoms. (The uncertainty in the $|E_b|/k_B$ values due to the laser line width of 200 MHz is $\approx 10$ mK.)

The beams from both 480 nm lasers are combined and made parallel using a polarizing beamsplitter cube, and half wave plates in each laser beam just before the beamsplitter cube allow the pulse energies in each beam to be varied independently. We calibrate the number of Rydberg atoms and ions created using the fluorescence depletion technique \cite{han09}. We monitor a fraction of the 780 nm fluorescence using a photomultiplier tube detector (PMT), and observe how much this is depleted when $5p_{3/2}$ atoms are either excited using the narrow bandwidth laser, or ionized using the Littman laser, when an SFI pulse is applied immediately afterwards. The SFI pulse removes Rydberg atoms or ions from the MOT so they no longer contribute to the cooling cycle. By measuring the resulting fall in the MOT atom population using the 780 nm fluorescence, we can calculate how many are excited to a Rydberg state, or are ionized. For some of the experiments, we reduced the $Q$-switch firing rate of both Nd:YAG lasers to 10 Hz or 6.67 Hz to reduce the steady state trap depletion and obtain higher Rydberg atom and ion densities. The laser beams are not focused into the MOT chamber, and the laser spot sizes are of order 3-4 mm, much larger than the radius of the cloud of cold atoms. We typically create plasmas with an initial ion number $N_{ion} = 5.0 \times 10^5$ (a maximum average initial ion density $\rho_{ion} \approx 3 \times 10^8$ cm$^{-3}$) and an initial number of Rydberg atoms typically in the range $N_{R} = (0.2 \, - \, 0.3) \times N_{ion}$. (The ion and Rydberg atom densities have an absolute uncertainty of a factor of approximately 2, and a relative uncertainty of 20-30\%.)

The UNPs created by the Littman laser are allowed to evolve in an environment in which the effects of stray external electric fields are minimized. The copper field meshes are stretched flat over stainless steel o-rings and are separated by 1.9 cm. We apply a small dc voltage to one of the field meshes (the other is grounded) so that the electric field in the interaction region is less than 10 mV/cm. (The interaction region is shielded in the perpendicular directions by two pairs of parallel metal plates spaced by approximately 10 cm.) Electrons or ions which exit through one of the field meshes are accelerated in a field of $\approx 10$ V/cm to microchannel plate detector (MCP). This field is created by biasing two meshes in front of the MCP which is itself inside a grounded metal enclosure so that field leakage is minimized. Our protocol for setting the dc voltage is to maximize the observed plasma lifetime when the electron TOF signal is being detected while maintaining a reasonable signal-to-noise ratio in the observed MCP signal. In these experiments, typical plasma lifetimes measured using the electron TOF signal are between 80 and 150 $\mu$s, while the typical duration of the ion TOF signal is such that all the ions reach the MCP within 250 $\mu$s of the lasers firing.

From the electron and ion TOF spectra, we extract the asymptotic plasma expansion velocity, $v_0$, a macroscopic parameter which depends on the initial electron and ion temperatures, plus heating caused by DIH and TBR, as well as heating or cooling caused by the added Rydberg atoms (see \cite{fore18} for a discussion of the utility and meaning of $T_{e,0}$). Practically, if we assume self-similar expansion of a Gaussian density profile of electrons and ions in the plasma, $v_0$ is related to the characteristic size of the plasma, $\sigma$, by $\sigma = \sqrt{\sigma^2_0 + v^2_0 \, t^2}$, where $t$ is the time since the plasma was created, and $\sigma_0$ is initial $e^{-1/2}$ radius of the plasma (this is assumed to be the same as that of the parent atoms in the MOT, though knowing the exact value of $\sigma_0$ is unnecessary to obtaining any of our results). The method we use to extract $v_0$ values from the Rb$^+$ ion TOF spectra is described in Ref. \cite{fore18}.

\begin{figure*}
\centerline{\resizebox{0.85\textwidth}{!}{\includegraphics{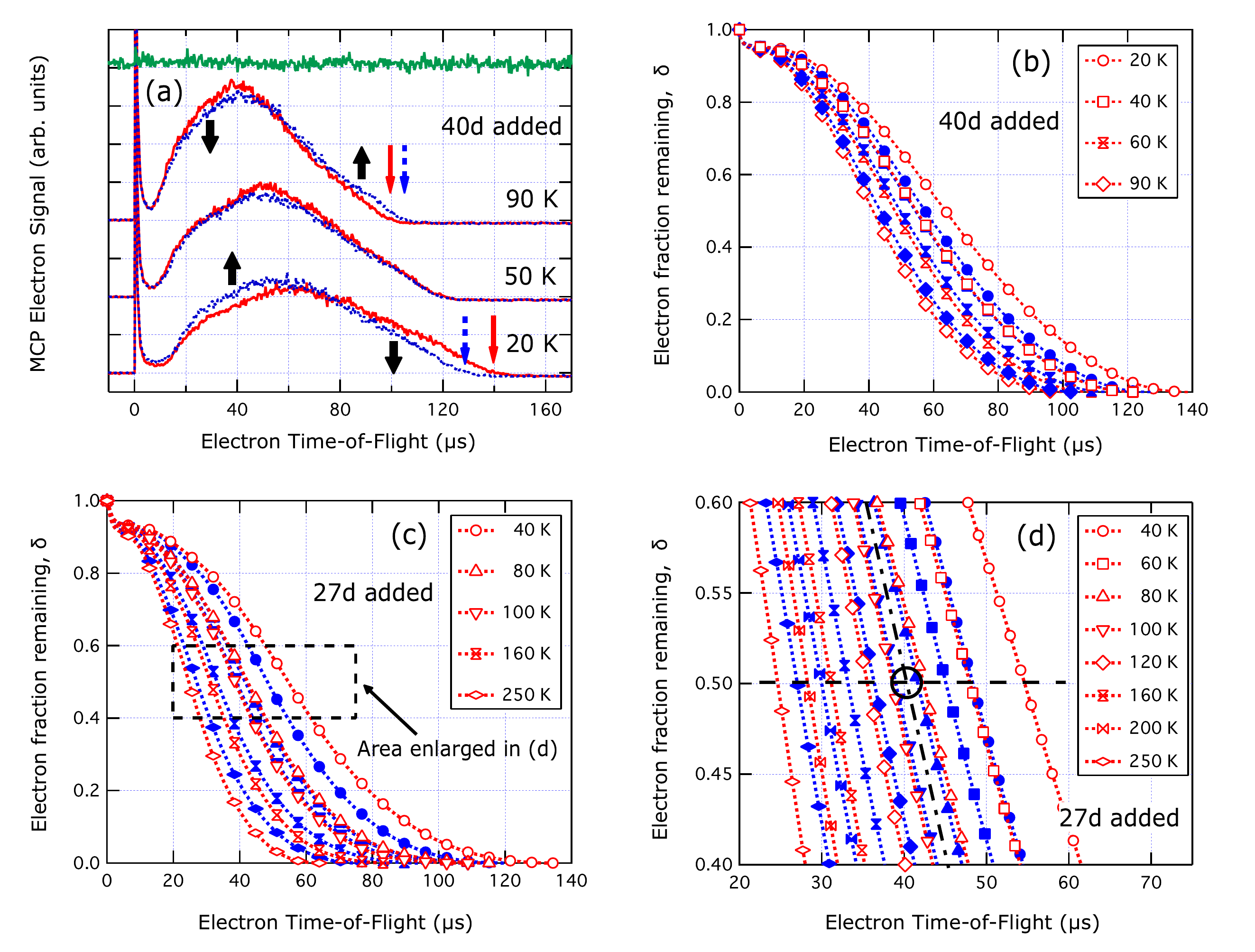}}}
\caption{Experimental electron TOF spectra of bare and embedded plasmas. (a) Electron signal detected by the MCP for UNPs with initial average ion density $\rho_{ion} = 4.4 \times 10^7$ cm$^{-3}$ and $T_{e,i} = 20$, 50, and 90 K with no Rydberg atoms added (solid red line) and with $40d$ atoms added with $N_R/N_{ion} = 0.28$ (blue dashed line). At top is the TOF signal with only the NBPL laser exciting the $40d$ state, multiplied by 10, showing that there is no spontaneous plasma formation from a sample of cold $40d$ atoms with density $1.2 \times 10^7$ cm$^{-3}$. The short fat black $\uparrow$ and $\downarrow$ arrows on the 20 K and 90 K TOF spectra represent the change (increase or decrease, respectively) in the electron signal at early and late times when Rydberg atoms are added. The red (solid) and blue (dashed) arrows represent the approximate lifetimes of the bare plasma, and the embedded plasma, respectively. (b) Data for UNPs embedded with $40d$ Rydberg atoms, plotted in terms of a $y$-coordinate which is the remaining electron fraction, $\delta$, defined using Eq. \ref{deltadef}. The red (open) symbols show the results for the bare plasma, while the blue (filled) symbols show the results for the plasma when Rydberg atoms are added, and the $T_{e,i}$ values are as shown in the legend. The uncertainty of the $\delta$ values at a given flight time is at the $\pm0.01$ level, approximately equivalent to the symbol size in each plot (see Sect. \ref{eTOF} for a discussion of how these uncertainties are obtained). (c) Plot of $\delta$ vs time for UNPs with different $T_{e,i}$ values when $27d_{5/2}$ atoms are added, for $\rho_{ion} = 2.1 \times 10^7$ cm$^{-3}$ and $N_R/N_{ion} = 0.19$. (d) Enlargement of the data in panel (c) in the region around $\delta = 0.5$ (the data set in panel (d) is more extensive than that in panel (c)). The black $- \cdot -$ line shows the approximate location of the crossover condition, where the addition of the Rydberg atoms does not affect the electron TOF signal and the horizontal black dashed line shows the condition $\delta = 0.5$. The black circle at the intersection of these lines is the crossover point, and $T_{CO}$ when $27d_{5/2}$ atoms are added to a UNP is the value of $T_{e,i}$ for a bare plasma that reaches $\delta = 0.5$ at a time of 40 $\mu$s. From these data, using interpolation as described in the text, we obtain $T_{CO} = 94 \pm 11$ K for a UNP with $27d_{5/2}$ atoms embedded in it.}
\label{eTOFspectra}
\end{figure*}

Unfortunately, there is no similar method that we are aware of by which absolute values of $v_0$ can be obtained from the electron TOF spectra. However, it is possible to obtain relative measurements of the expansion velocity from the electron TOF spectra, and we used these to obtain a set of independent measurements of the crossover condition. Twedt and Rolston \cite{twed10} studied electron evaporation from a UNP in a dc electric field, and found that the remaining electron fraction, $\delta$, depends in a simple manner on the parameter 
\begin{equation}
\alpha = \frac{E_{ext}\, \epsilon_0 \, \sigma^2}{e \, N_{ion}}, \label{alpha}
\end{equation}
where $E_{ext}$ is the applied electric field. We have used Twedt's code and obtained the heuristic relationship $\delta = 1 - \exp(-223 \, \alpha)$, which reproduces the data in Ref. \cite{twed10} within a few percent for $\delta \ge 0.1$. We use our electron TOF spectra to obtain the value of $\alpha$ for a particular value of $\delta$, say $\delta = 0.5$ (i.e., $\alpha = 3.11 \times 10^{-3}$). Then, using Eq. \ref{alpha}, and assuming that $v_0 \, t \gg \sigma_0$, we can find $v_0$ using
\begin{equation}
v^2_0 = \biggl (\frac{e \, N_{ion}}{\epsilon_0 \, E_{ext}} \biggr ) \, \frac{\alpha(\delta)}{t^2_\delta}, \label{delta}
\end{equation}  
where $t_\delta$ is the time at which the condition $\alpha(\delta = 0.5)$ occurs. Since it is hard to measure $E_{ext}$ and $N_{ion}$ without introducing significant additional error, we simply use the quantity $\sqrt{\alpha(\delta)}/t_\delta$ as a proxy for $v_0$. If, when we add Rydberg atoms, $\sqrt{\alpha(\delta)}/t_\delta$ increases, we say the plasma has been heated; if it decreases, then the plasma has cooled. 

There are obvious shortcomings to this procedure. First, for embedded plasmas in which a significant fraction of the Rydberg atoms ionize, $N_{ion}$ will be different to that for a plasma which did not have Rydberg atoms added to it. Second, the plasma may not have achieved a constant value of $v_0$ at the time when $\delta = 0.5$, especially in the situation where Rydbergs have been added to the plasma. Thirdly, the time at which $\delta = 0.5$ changes as $T_{e,i}$ is changed. If $v_0$ has truly reached its asymptotic value, this should not matter, but if $v_0$ is still changing, there is a potential systematic error from this source. Finally, Ref. \cite{twed10} assumes that the electrons are a zero-temperature fluid, and does not account for evaporation of electrons from the plasma that happens at times before the ion well potential plus the external field would allow spilling to occur. The possible impacts of this simplification are discussed in Sect. \ref{disc}.

\section{{RESULTS}\label{res}}
\subsection{{Crossover condition: Electron TOF spectra}\label{eTOF}} 
Examples of electron TOF spectra for UNPs created with different $T_{e,i}$ values when Rydberg atoms in the $27d_{5/2}$ and $40d$ states are embedded are shown in Fig. \ref{eTOFspectra}. In Fig. \ref{eTOFspectra}(a), we show plasmas created with $T_{e,i} = 20$, 50, and 90 K, both with and without the addition of $40d$ atoms (binding energy magnitude $|E_b|/k_B = 106$ K). The number of ions is $N_{ion} = 6 \times 10^5$ (average ion density $\rho_{ion} = 4.4 \times 10^7$ cm$^{-3}$) and initial ratio $N_{R}/N_{ion} = 0.28$. Each of the TOF spectra shown in Fig. \ref{eTOFspectra} have a time resolution of 64 ns, and are averages over eight laser shots. Additionally, the spectra were subsequently smoothed by averaging over five adjacent time points. The averaged TOF spectra have a root mean square noise amplitude of $\sim 1$\% of the signal size at the maximum of the ``hump'' at 40-50 $\mu$s flight time (the maximum noise variation of the signal is at the $\pm2$\% level).

As can be seen, adding $40d$ atoms to a $T_{e,i} = 20$ K plasma increases the electron evaporation signal relative to the UNP with no embedded Rydberg atoms (which we term a ``bare'' plasma) at early evolution times, and the plasma lifetime (i.e., the time for all the electrons to evaporate from the UNP) is shorter than that of the bare plasma by approximately 10 $\mu$s. (The changes in the electron evaporation signal due to the addition of the Rydberg atoms is significantly greater than the noise level of the averaged TOF signals.) However, adding $40d$ atoms to the $T_{e,i} = 90$ K UNP reduces the early time electron evaporation signal, and the plasma lifetime is slightly longer than for the bare plasma. On the other hand, adding $40d$ atoms to the $T_{e,i} = 50$ K UNP has an almost negligible effect on the plasma lifetime and electron evaporation rate. In other words, $40d$ atoms accelerate the evolution of the 20 K plasma, leave the 50 K plasma almost unchanged, but slow the evolution of the 90 K plasma. (The upper trace in Fig. \ref{eTOFspectra}(a) is the electron signal from the $40d$ atoms when there is no UNP present multiplied by a factor of 10, showing that there is clearly no evidence of spontaneous plasma formation.) 

The effect of adding $40d$ atoms to the plasma is more clearly seen in Fig. \ref{eTOFspectra}(b), where the remaining electron fraction, $\delta$, is plotted versus time \cite{twed10}. Experimentally, $\delta$ at time $t$ is obtained using the equation
\begin{equation}
\delta(t) = 1 - \frac{\int^t_0 \, S_e(t) \, dt}{\int^\infty_0 \, S_e(t) \, dt}, \label{deltadef}
\end{equation} 
where $S_e(t)$ is the electron TOF signal detected by the MCP, i.e., data like that shown in Fig. \ref{eTOFspectra}(a). For $T_{e,i} = 20$ and 40 K, the time for the UNP to reach a particular value of $\delta$ ($\delta = 0.5$, say) decreases when Rydberg atoms are added, while for $T_{e,i} = 60$ and 90 K, the time needed to reach a given $\delta$ increases when Rydberg atoms are added. Hence, the data shown in Figs. \ref{eTOFspectra}(a) and \ref{eTOFspectra}(b) show that the crossover $T_{e,i}$ value when $|E_b|/k_B = 106$ K is approximately 50 K. The effect of adding $27d_{5/2}$ atoms ($|E_b|/k_B = 240$ K) to UNPs with various different $T_{e,i}$ values is shown in Figs. \ref{eTOFspectra}(c) and \ref{eTOFspectra}(d). For plasmas that are embedded with $27d_{5/2}$ atoms, the crossover condition is approximately $T_{CO} = T_{e,i} = 90$ K. 

The estimated vertical uncertainties of the $\delta$ versus time curves in Figs. \ref{eTOFspectra}(b)-\ref{eTOFspectra}(c) are less than $\pm 0.01$, which approximately corresponds to the size of the line symbols in Figs. \ref{eTOFspectra}(b) and \ref{eTOFspectra}(c). This estimate was obtained by looking at the $\delta$ versus time curves for bare plasmas with $T_{e,i} = 40$ K and 100 K taken over the course of one day. The standard deviation of the $\delta$ values at a particular time of flight for a given $T_{e,i}$ was found to be $\approx 0.01$ for data taken on one day, although it was somewhat larger, $\approx 0.03$, when the data taken over the course of a week are considered. Such variations arise due to different plasma and Rydberg densities caused by different MOT atom densities, and different laser pulse energies and line widths. These effects determine the estimated $\pm 0.03$ long term uncertainty in $\delta$. However, as described below, we always obtain TOF spectra in pairs, one of a bare plasma and one of an embedded plasma with the same $T_{e,i}$, at time intervals of a minute or less. Hence, the actual uncertainties which affect our ability to distinguish the effects of adding Rydberg atoms to UNPs in the $\delta$ versus time curves are of the same order as the daily variation, $\pm 0.01$. 

We obtained electron TOF spectra like those shown in Fig. \ref{eTOFspectra} for 15 different Rydberg states with $|E_b|/k_B$ values in the range 46 K ($60d$) to 308 K ($24d_{5/2}$). Each Rydberg state was embedded into plasmas with a range of $T_{e,i}$ values. At least six different temperatures were used for each Rydberg state, and sometimes as many as nine temperatures were used, and we obtained duplicate data sets for two different Rydberg states taken on different days to get a sense of day-to-day variations in the data. For each $T_{e,i}$, $E_b$ combination, we obtained an electron TOF spectrum for a plasma with no Rydberg atoms embedded, followed immediately afterwards by a spectrum from an embedded plasma. To find the crossover $T_{e,i}$ for the $|E_b|$ of the added Rydberg atoms, we used the two TOF spectra at each $T_{e,i}$ (one for the UNP with $N_R$ Rydberg atoms embedded, the other for the bare plasma) to find a proxy for the change in the effective initial electron temperature of the UNP when atoms are added, $\Delta T_{N_R}$. Specifically, using Eq. \ref{vZero}
\begin{equation}
\Delta T_{N_R} = \frac{m_{ion}}{k_B} \, [v^2_0 ({N_R}) - v^2_0 ({0})], \label{deltaT}
\end{equation}
where $v_0 (N_R)$ is the plasma expansion velocity when $N_R$ Rydberg atoms are added, and $v_0 (0)$ is the corresponding velocity for the bare plasma ($N_R=0$). As described in Sect. \ref{exp}, we cannot use the electron TOF spectra to find the $v_0$. Instead, we used the proxies $v^\ast_0 = \sqrt{\alpha(\delta=0.5)}/t_{0.5} = 5.58 \times 10^{-2}/t_{0.5}$, and ${T}^\ast_{e,0} = {v^\ast_0}^2$, where $t_{0.5}$ is the time at which $\delta=0.5$. The proxy for the change in the effective initial electron temperature is thus
\begin{equation} %
\Delta T^\ast_{N_R} = 3.11 \times 10^{-3} \, \biggl (\frac{1}{[t_{0.5} (N_R)]^2} - \frac{1}{[t_{0.5} (0)]^2} \biggr ) , \label{deltaTproxy}
\end{equation} 
and we performed a regression of $\Delta T^\ast_{N_R}$ versus $T_{e,i}$ to find the $T_{e,i}$ value for which $\Delta T^\ast_{N_R} = 0$. This $T_{e,i}$ value is the UNP crossover electron temperature, $T_{CO}$, for the specific $|E_b|$ of the added Rydberg atoms. The uncertainties in the $\delta$ values ($\pm 0.01$) are approximately five times smaller than the change in $\delta$ for a  bare plasma near flight times corresponding to $\delta = 0.5$ when Rydberg atoms are added to the plasma for which $T_{e,i}$ is significantly different to $T_{CO}$. When $T_{e,i}$ is closer to $T_{CO}$, the uncertainties in $\delta$ become larger than the change when Rydberg atoms are added. The variations in experimental conditions which determine the uncertainties in $\delta$ are also manifested in the scatter of the data points, and consequently determine uncertainties of the $T_{CO}$ values obtained from the regressions. We used the regression uncertainties, along with the uncertainties in the $T_{e,i}$ values from the Littman laser calibration, to obtain the final uncertainties in the crossover temperatures.


\subsection{{Crossover condition: Ion TOF spectra}\label{iTOFx}} 

\begin{figure}
\centerline{\resizebox{0.50\textwidth}{!}{\includegraphics{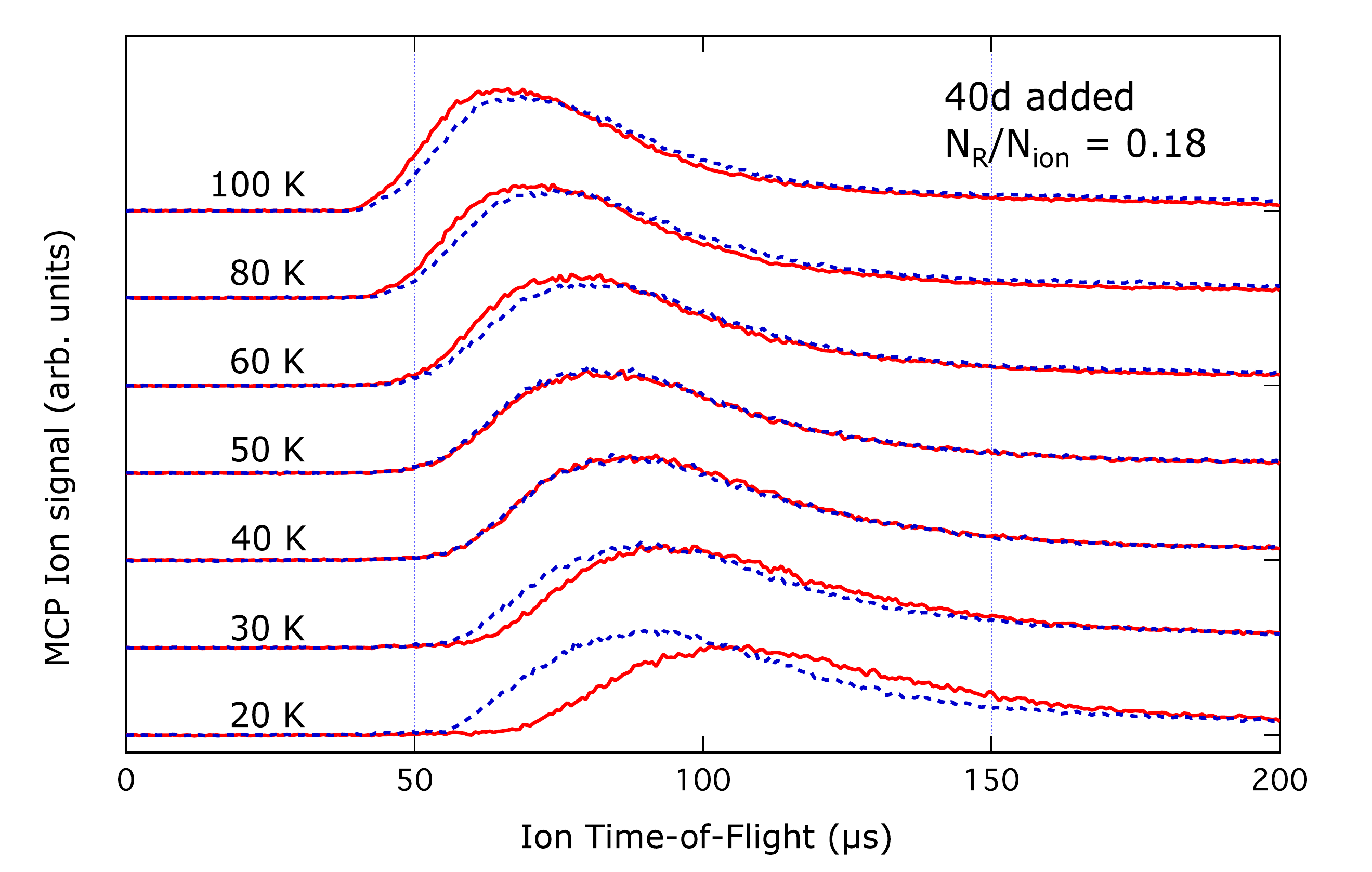}}}
\caption{Experimental ion TOF spectra for UNPs embedded with $40d_{5/2}$ atoms ($|E_b|/k_B = 106$ K). The red solid lines are the TOF signals for the bare (non-embedded) plasmas, and the blue dashed lines are for the embedded plasmas. The $T_{e,i}$ values defined using Eq. \ref{Teinit} are shown on each trace (the uncertainty in $T_{e,i}$ is $\pm5$ K). As with the electron TOF spectra shown in Fig. \ref{eTOFspectra}(b), the crossover condition is somewhere in the range $T_{e,i} = 40$ - 50 K. For these data, $N_{ion} = 4 \times 10^5$ (initial average $\rho_{ion} = 4.5 \times 10^7$ cm$^{-3}$), and $N_R/N_{ion} = 0.18$. These ion TOF spectra were averaged over 128 laser shots, and the TOF spectra were also subsequently smoothed by averaging them over five adjacent time points (the time resolution of the raw data was 128 ns).}
\label{iTOFspectra}
\end{figure}

We also studied the crossover condition using the ion TOF spectra. Specifically, we looked at UNPs with $24d_{5/2}$, $27d_{5/2}$, $32d_{5/2}$, $40d$, and $60d$ Rydberg atoms embedded in them. Spectra obtained when the $40d$ state ($|E_b|/k_B = 106$ K) was embedded are shown in Fig. \ref{iTOFspectra}. As can be seen, in UNPs with $T_{e,i} \ge 60$ K, the ions take slightly longer to reach the MCP when Rydberg atoms are embedded, indicating that the addition of Rydberg atoms to the plasma decreases $v_0$. On the other hand, when $T_{e,i} \le 30$ K, $v_0$ increases when Rydberg atoms are embedded, and the crossover condition is in the range $T_{e,i} = 40$ - 50 K. This is the same range for the crossover condition given by the electron TOF spectra in Fig. \ref{eTOFspectra}(b).

We obtained five crossover values from the ion TOF spectra, for the $24d_{5/2}$, $27d_{5/2}$, $32d_{5/2}$, $40d$, and $60d$ states. These data were obtained with $N_R/N_{ion}$ values in the range 0.2 - 0.6. Using the approach described in Ref. \cite{fore18}, we were able to use the ion TOF spectra to obtain explicit $v_0$ values, and we use these to regress $\Delta T$ as defined in Eq. \ref{deltaT} versus $T_{e,i}$ (as with the electron spectra, the ion TOF spectra were obtained in pairs for a particular $T_{e,i}$ value, one for a plasma with Rydberg atoms added, and one with no added atoms). The $v_0$ values obtained by this method are subject to uncertainties which have the same origin as those which affect the electron TOF spectra, namely, shot-to-shot variations in the ion and Rydberg atom densities. However, the ion TOF spectra were averaged over 128 laser shots, and the TOF spectra were also subsequently smoothed by averaging them over five adjacent time points (the time resolution of the raw data was 128 ns). The technique of taking spectra in pairs and comparing $v_0$ values from ion signals which were acquired within one or two minutes of each other minimizes the impact of the shot-to-shot variations in the experimental conditions. The effect of such variations are manifested in the uncertainties of the values of $T_{CO}$ obtained from the regression.  

\subsection{{Crossover condition: Summary of results}\label{xoverSum}}
\begin{figure*}
\centerline{\resizebox{0.80\textwidth}{!}{\includegraphics{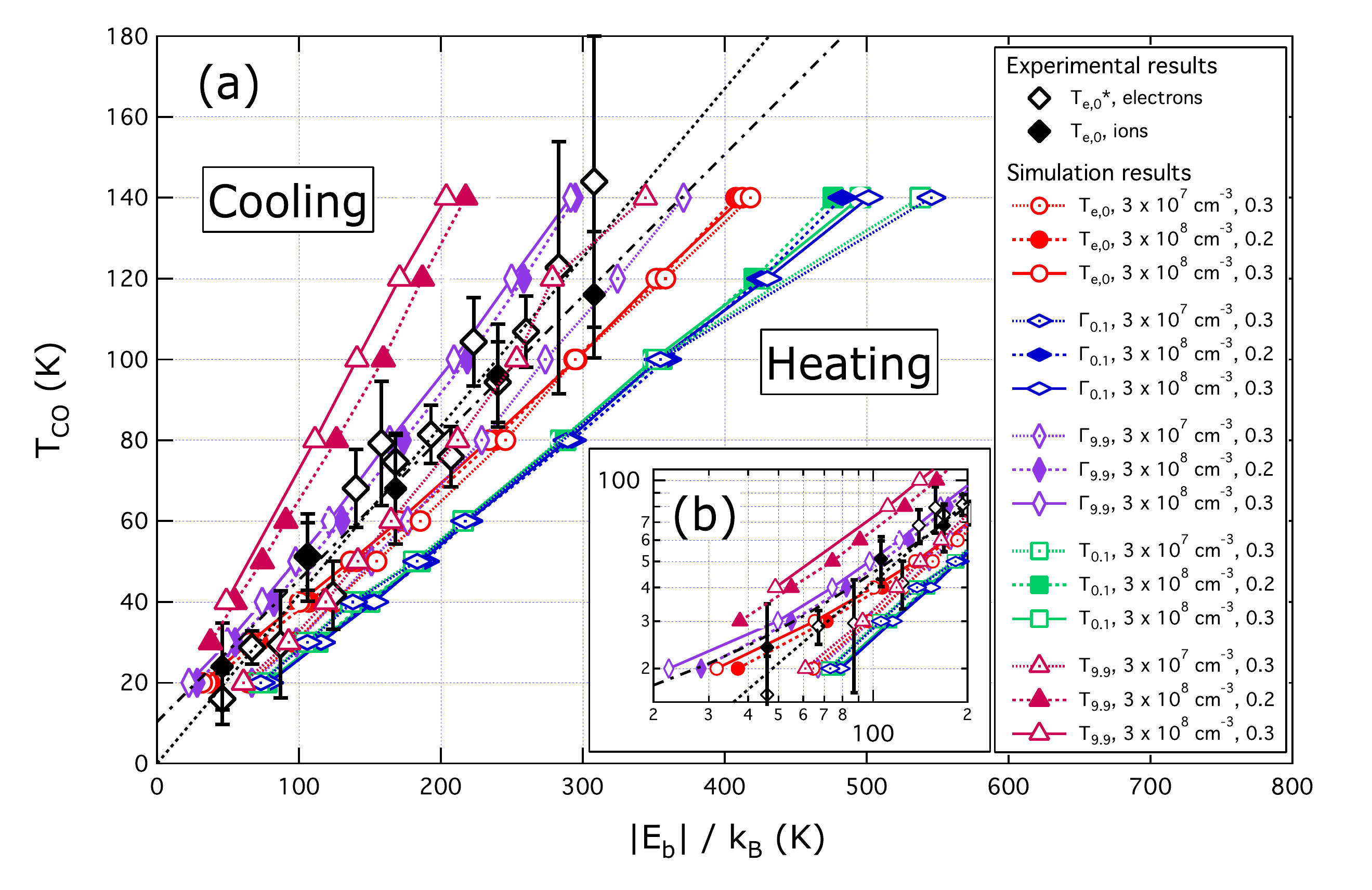}}}
\caption{(a) Experimental and numerical results for the crossover electron temperature, $T_{CO}$, versus the magnitude of the binding energy of the added Rydberg atoms in units of K, $|E_b|/k_B$. Inset (b) is an enlargement of the region $|E_b|/k_B \le 200$ K, $T_{CO} \le 110$ K on a log-log scale. In both graphs, the experimental results are shown for electrons ($\lozenge$) and ions ($\blacklozenge$) when the effective initial electron temperature of the plasma, $T_{e,0}$, is the marker parameter. The average ion densities for these data are in the range $3 - 7 \times 10^7$ cm$^{-3}$, and $\sigma_0 \approx 550$ $\mu$m. The black dotted line is a weighted fit to the experimental electron data, and the black dash-dot-dashed line is a fit to the experimental ion data. The simulation results are for the crossover conditions for the marker parameters electron temperature and electron Coulomb coupling parameter at 100 ns, $T_{0.1}$ and $\Gamma_{0.1}$ respectively, the electron temperature and coupling parameter at 9.9 $\mu$s, $T_{9.9}$ and $\Gamma_{9.9}$ respectively, as well as the value of $T_{e,0}$ found from the average expansion velocity over 9.9 to 19.9 $\mu$s of evolution time. The simulations were run with ion densities in the range $0.3 - 3 \times 10^8$ cm$^{-3}$ and $\sigma_0 = 354 \ \mu$m. The legend gives the average ion density, $\rho_{ion}$, and the ratio of Rydberg atoms to ions, $N_R/N_{ion} = 0.2$ or 0.3, for the simulations. The experimental vertical error bars are as shown, while the horizontal error bars are negligible (see Sects. \ref{res} A, B, and C for an explanation of how the error bars were obtained). The simulation error bars are not shown: The $T_{CO}$ values are known exactly since this was set as an initial condition in each simulation, while the uncertainty in the $|E_b|/k_B$ values is $\pm 2.5$ K for $|E_b|/k_B = 100$ K and $\pm 25$ K for $|E_b|/k_B = 500$ K, equivalent to one half of the adjacent Rydberg state energy spacing. The parameter space to the left of a line for a given marker on the graph corresponds to cooling of the plasma as measured using that marker (i.e., the marker reaches a particular value at a longer evolution time when Rydberg atoms are added to the plasma than for the bare plasma) while to the right of the line corresponds to heating. }
\label{tCO}
\end{figure*}

The 15 $T_{CO}$ values obtained from the electron TOF spectra and the five $T_{CO}$ values obtained from the ion TOF spectra are plotted versus $|E_b|$ in Fig. \ref{tCO}, and are discussed in Sect. \ref{iXMod}. For each experimental data point, the vertical error bar is found by adding the regression uncertainty and the $\pm 5$ K uncertainty of the experimental $T_{e,i}$ values in quadrature. There is essentially no uncertainty in the $|E_b|/k_B$ values since the NBPL was tuned to resonance as verified from the SFI signal (the laser line width of $\approx 200$ MHz corresponds to a 10 mK uncertainty in $|E_b|/k_B$).

The crossover condition obtained from the electron data shown in Fig. \ref{tCO} can be expressed as $|E_b| = (2.4 \pm 0.2) \times k_B \, T_{CO}$ (intercept$/k_B$ = $0 \pm 10$ K). For the ion data, the $T_{e,0}$ crossover condition is $|E_b| = (2.9 \pm 0.5) \times k_B \, T_{CO}$ (intercept$/k_B$ = $-30 \pm 30$ K), but if the $y$-intercept is constrained to be zero, the result is $|E_b| = (2.5 \pm 0.2) \times k_B \, T_{CO}$. As discussed in Sect. \ref{iXMod}, the $T_{CO}$ values obtained from the ion and electron TOF spectra likely correspond to slightly different measures of plasma behavior (``markers''). While the crossover values obtained from the ion spectra closely correspond to the $T_{e,0}$ marker, that from the electron TOF spectra may be more closely related a different marker, the Coulomb coupling parameter for the electrons. It therefore does not make sense to average these crossover behaviors together, and perhaps the most accurate summary that covers both the ion and electron experimental results is to express the crossover condition as $|E_b| = 2.7 (\pm 0.5) \times k_B \, T_{CO}$.

\subsection{{Amount of heating or cooling}\label{ieTOFheat}} 

\begin{figure}
\centerline{\resizebox{0.45\textwidth}{!}{\includegraphics{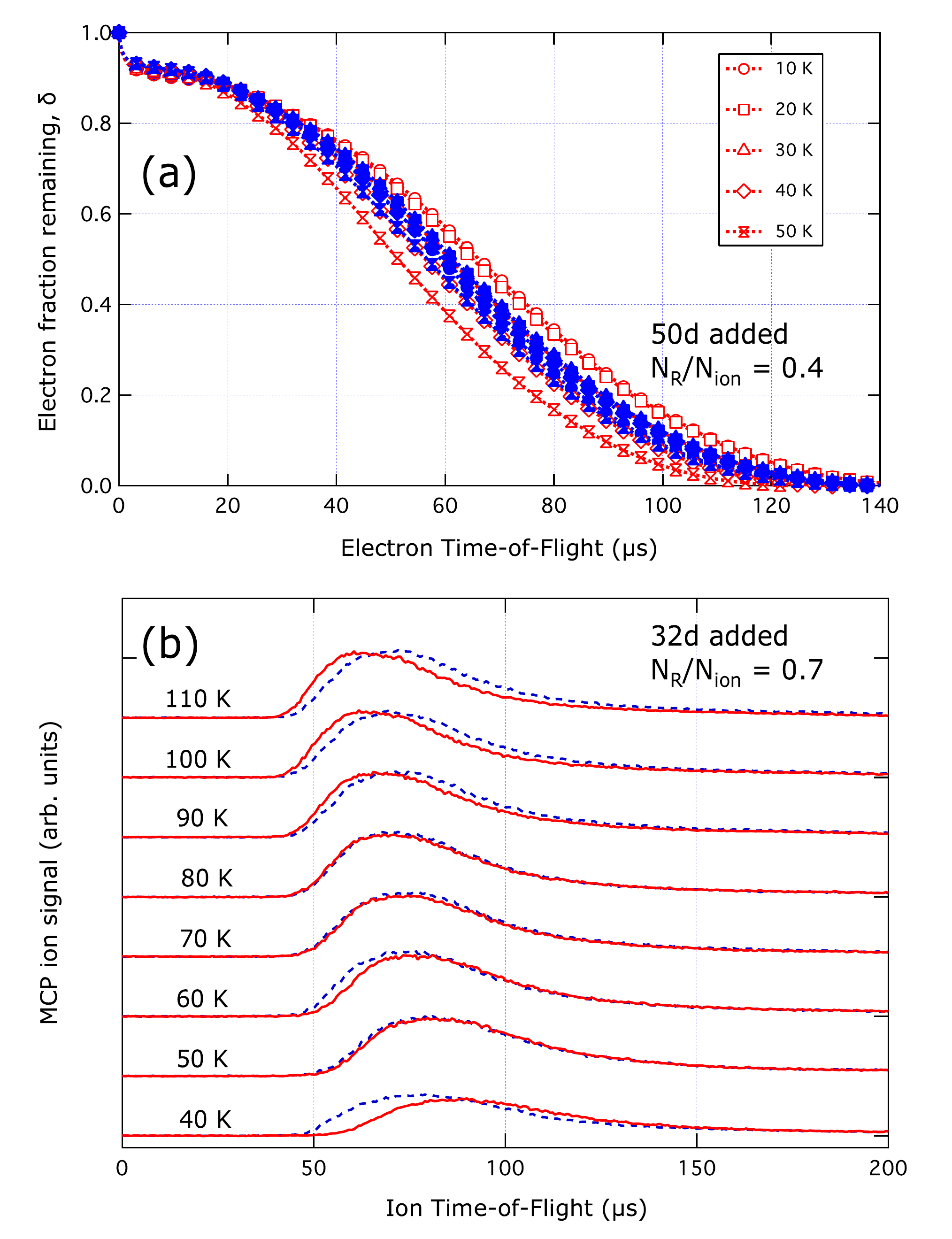}}}
\caption{(a) Experimental electron TOF spectra plotted in terms of the remaining electron fraction when $50d$ Rydberg atoms are added to UNPs with various $T_{e,i}$ values as shown in the legend, when $N_R/N_{ion} = 0.4$ and $\rho_{ion} = 4.0 \times 10^7$ cm$^{-3}$. The red (open) symbols show the results for the bare plasma, while the blue (filled) symbols show the results for the plasma when Rydberg atoms are added. (b) Ion TOF spectra when $32d$ Rydberg atoms are added to UNPs with $T_{e,i}$ values as shown, for $N_R/N_{ion} = 0.7$ and $\rho_{ion} = 4.4 \times 10^7$ cm$^{-3}$. The TOF spectra for the bare plasmas are shown by a solid red line, the spectra for the UNPs embedded with Rydberg atoms are shown by the blue dashed lines. As can be seen in both the electron and the ion TOF spectra when Rydberg atoms are embedded, there is hardly any variation in the spectra with $T_{e,i}$, indicating that the plasma behavior is almost completely controlled by the added Rydberg atoms.}
\label{iAndeTOFspectra}
\end{figure}

In the experiments where we acquired electron and ion TOF spectra to find values for the crossover $|E_b|$ values for a given $T_{e,i}$, we attempted to keep the ratio $N_R/N_{ion} \lesssim 0.3$, in order to minimize the possibility that the embedded Rydberg atoms would form a plasma spontaneously and independently of the co-created plasma. Unfortunately, given the sensitivity of the 780 nm fluorescence to MOT cooling and repump laser frequency drifts (which impact our ability to measure $N_R/N_{ion}$ accurately), and frequency drifts of the 960 nm ECDL used to embed the Rydberg atoms, it was not always possible to maintain $N_R/N_{ion} \lesssim 0.3$. Consequently, we sometimes obtained data at larger $N_R/N_{ion}$ values, though we always checked that the electron and ion TOF spectra showed no evidence of spontaneous evolution of the Rydberg atoms to plasma when the Littman photoionizing laser beam was blocked. Inspection of TOF spectra obtained with $N_R/N_{ion} > 0.3$ showed that these UNPs evolved in a manner that was significantly decoupled from the $T_{e,i}$ value set by the Littman laser frequency, and the plasma expansion seemed to depend almost exclusively on the embedded Rydberg atoms. Examples of this behavior are shown in Fig. \ref{iAndeTOFspectra}. In Fig. \ref{iAndeTOFspectra}(a), the $\delta$ versus time signatures obtained from electron TOF spectra with $T_{e,i}$ values in the range 10 - 50 K when the $50d$ state ($|E_b|/k_B = 67$ K) is embedded with $N_R/N_{ion} = 0.4$ are shown. All of these plasmas evolve such that $\delta(t)$ is very close to that for a plasma with no added Rydberg atoms and $T_{e,i} = 25$ K. This is close to the crossover $T_{e,i}$ we found as described in Sect. \ref{eTOF} for a UNP embedded with $50d$ atoms, which was 29 $\pm$ 4 K. Similar behavior is apparent in the ion TOF spectra, as can be seen in Fig. \ref{iAndeTOFspectra}(b), which shows plasmas embedded with $32d_{5/2}$ atoms ($|E_b|/k_B = 168$ K) at $N_R/N_{ion} = 0.7$. As can be seen, the TOF spectra for plasmas with embedded Rydberg atoms are almost identical, and are all similar to the UNPs with no embedded Rydberg atoms with $T_{e,i} = 70$ - 80 K ($T_{CO}$ for $32d_{5/2}$ embedded plasmas was 75 $\pm$ 7 K).

The behavior of the electron and ion TOF spectra when Rydberg atoms are embedded in UNPs with high $N_R/N_{ion}$ values are clearly different from those seen at lower relative Rydberg atom densities, which are shown in Figs. \ref{eTOFspectra} and \ref{iTOFspectra}. Specifically, at lower Rydberg atom densities, the electron and ion TOF spectra depend mostly on the $T_{e,i}$ set by the photoionization laser, though the plasma evolution is accelerated by adding Rydberg atoms when $T_{e,i}$ is less than the $T_{CO}$ value for the $|E_b|$ of the added atoms, and slowed down when $T_{e,i} > T_{CO}$. On the other hand, plasmas embedded with Rydberg atoms at higher relative density evolve in a manner which is identical to the signature of an non embedded plasma with $T_{e,i}$ equal to the crossover value determined by the atom binding energy.

We investigated how the Rydberg atom embedded UNPs behave between these two limiting behaviors. Specifically, we looked at ion TOF spectra of UNPs embedded with $27d_{5/2}$ atoms, and we varied $N_R/N_{ion}$ from zero (no embedded Rydberg atoms) up to 0.53. The crossover $T_{e,i}$ for UNPs embedded with $27d_{5/2}$ atoms was found to be 94 $\pm$ 11 K (as described in Sect. \ref{iTOFx}), and we used $T_{e,i}$ values of 40, 50, 60, 80, 120, 140, 170, and 200 K. For each of seven different $N_R/N_{ion}$ values, we obtained ion TOF spectra when $T_{e,i}$ was set to these eight values. These spectra were then fitted using the technique described in Ref. \cite{fore18} to obtain values for $v_0$, the plasma asymptotic expansion velocity. Then, using Eq. \ref{vZero}, we obtained values for the effective initial electron temperature of the embedded plasmas, $T_{e,0}$, by assuming that $T_{ion,0}$ is negligible. 

\begin{figure}
\centerline{\resizebox{0.45\textwidth}{!}{\includegraphics{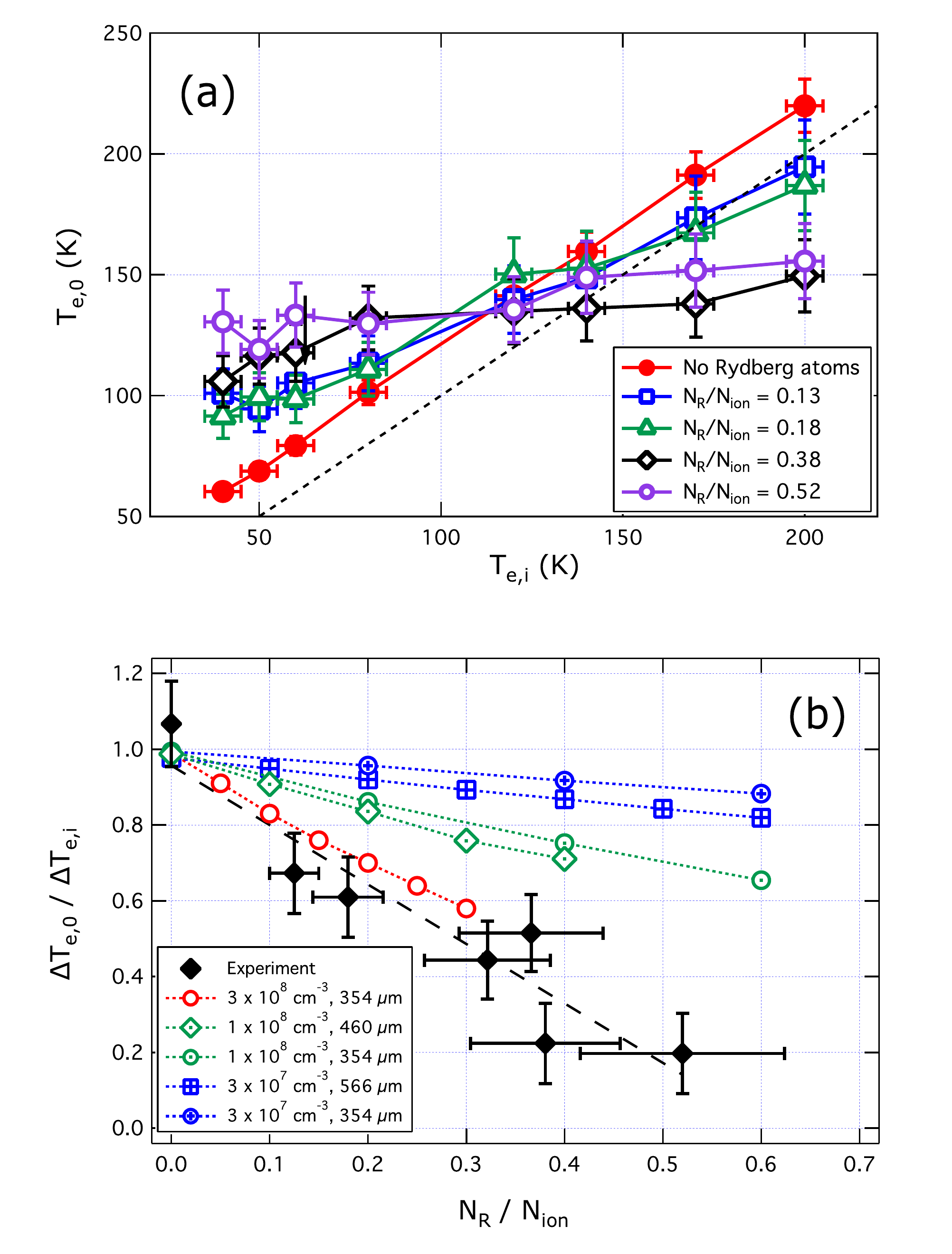}}}
\caption{(a) The effect of varying Rydberg atom density (relative to ion density) when $27d_{5/2}$ atoms are added to UNPs with $T_{e,i}$ values as shown. The average ion density for these data is $(2.8 \pm 0.7) \times 10^8$ cm$^{-3}$, and $\sigma_0 = 560 \pm 40$ $\mu$m. The $T_{e,i}$ values are set by the frequency of the photoionization laser, and $T_{e,0}$ is found using Eq. \ref{vZero} and the $v_0$ values obtained from fitting the ion TOF spectra (we assume $T_{ion,0} = 0$). The black dashed line corresponds to $T_{e,0} = T_{e,i}$, and the fact that the results with $N_R = 0$ are consistently 20 K above this line indicates a likely systematic effect in the measurement technique. (b) Slope of $T_{e,0}$ vs $T_{e,i}$ found from straight line fits of the data in (a) vs $N_R/N_{ion}$. The black diamonds are the experimental data (i.e., those in panel (a), plus two additional data sets). The other symbols, connected by the dashed lines, are results of simulations described in Sect. \ref{iHeatMod} with initial average ion density and $\sigma_0$ values as shown. The black dashed line is a fit to the experimental data with slope $= -1.6 \pm 0.3$ and intercept $= 0.96 \pm 0.08$. The vertical and horizontal error bars for the experimental data in panels (a) and (b) are discussed in Sect. \ref{ieTOFheat}. The uncertainties of the simulation data in panel (b) are discussed in Sect. \ref{iHeatMod}.}
\label{heating}
\end{figure}

The results of this analysis are shown in Fig. \ref{heating}(a), for five different $N_R/N_{ion}$ values. As can be seen, the amount of cooling (that is, how much $T_{e,0}$  lies below the value of a bare plasma with the same $T_{e,i}$) when $T_{e,i}$ is above the crossover, and the amount of heating when $T_{e,i}$ is below the crossover value, are both approximately proportional to $T_{e,i} - T_{CO}$ and to $N_R/N_{ion}$:
\begin{equation}
T_{e,0}(N_R) - T_{e,0}(N_R=0) \propto -\frac{N_R}{N_{ion}} \, (T_{e,i} - T_{CO}). \label{expAmtHeat}
\end{equation}
As we have described in Ref. \cite{fore18}, there are nuances to the experimental technique used to find $v_0$ from the ion TOF spectra that limited the precision of the experimental $T_{e,0}$ values to $\sim10$\%. This value is found from the statistical variation of the expansion velocities found by fitting the ion TOF spectra from plasmas created under nominally identical conditions. The vertical error bars in Fig. \ref{heating}(a) are $\pm$10\% for $N_R/N_{ion} \ne 0$ and $\pm$5\% for the data points where $N_R/N_{ion} = 0$ (we obtained many more TOF spectra with $N_R/N_{ion} = 0$ than for embedded plasmas, giving better statistics for the $T_{e,0}$ values for bare plasmas). The horizontal error bars are $\pm 5$ K, which is determined by the precision of the frequency calibration of the Littman laser. 

The experiments for the present paper were analyzed as described in Ref. \cite{fore18}, and should be free of systematic effects that limit the accuracy of the results. (These effects, and their impact on the accuracy and precision of the experiments, are extensively described in Ref. \cite{fore18}.) Nevertheless, as can be seen in Fig. \ref{heating}(a), the data for $N_R/N_{ion}=0$ exhibit a systematic offset of approximately 20 K. The $N_R/N_{ion}=0$ data should fall on the line $T_{e,0} = T_{e,i}$ since TBR heating is negligible for these data (the UNP density is low and $T_{e,i} \ge 40$ K). The 20 K offset is likely caused by an inaccurate value for the time offset when the ion TOF spectra are fitted to obtain $v_0$. However, the fact that the $N_R/N_{ion}=0$ data are offset by the same amount strongly suggests that the systematic effect is independent of $T_{e,i}$. (It should be pointed out that the values of $T_{e,i}$ at the crossover condition found in Sects. \ref{eTOF} and \ref{iTOFx} are not subject to uncertainty from the 20 K offset since $T_{e,i}$ at the crossover is found from the ionizing laser frequency when $T_{e,0}$ is unchanged by the addition of Rydberg atoms.) We therefore fitted the data shown in Fig. \ref{heating}(a) assuming a linear dependence of $T_{e,0}$ on $T_{e,i}$ to obtain the slopes, $\Delta T_{e,0}/\Delta T_{e,i}$. These slopes are shown (versus $N_R/N_{ion}$) in Fig. \ref{heating}(b), along with the results of the numerical simulations described in Sect. \ref{iHeatMod}. The vertical error bars for the experimental data in Fig. \ref{heating}(b) are determined by the vertical and horizontal error bars in Fig. \ref{heating}(a) and the uncertainties of the slope values obtained from fits to the data in Fig. \ref{heating}(a). The horizontal error bars are estimated as $\pm20$\%, which is determined by the accuracy of our technique for measuring the ion and Rydberg atoms densities described in Sect. \ref{exp}. 

\section{{NUMERICAL ANALYSIS}\label{anal}}
As described above, we have measured, using both electron and ion TOF spectra, values of the crossover temperature, $T_{CO}$, when UNPs are embedded with Rydberg atoms with binding energy $E_b$. Additionally, we have measured the amount of heating or cooling which occurs in UNPs when Rydberg atoms are embedded, as functions of the Rydberg atom density and $T_{e,i}$. We describe here the numerical modeling we have carried out to understand these experimental results and gain an intuitive understanding of their underlying plasma dynamics.

\subsection{{Numerical modeling approach}\label{mod}} 

We have modeled the interaction between the UNP and the Rydberg atoms using Monte-Carlo simulations described in Ref. \cite{robx03}. The programs, which are based on the work of Mansbach and Keck \cite{mans69}, have been used as described in Ref. \cite{fore18}. Specifically, the simulations model the evolution of a UNP with specified initial electron temperature, average ion density, and initial radius, $\sigma_0$, as it interacts with a reservoir of embedded Rydberg atoms with average density $(N_R/N_{ion}) \, \rho_{ion}$ in an $nd$ state. The atoms are embedded at time $t = 0$, i.e., we assume that the Rydberg reservoir is created at the same instant as the UNP, and the simulations were run to a final evolution time of 20 $\mu$s. We ran two kinds of simulations. First, we found the crossover electron temperatures $T_{CO}$ of embedded plasmas using several different markers (measures of plasma behavior) by comparing against the same markers in bare plasmas with a specific $T_{e,i}$ using a variational approach in $|E_b|$. For each $T_{e,i}$, simulations in which between five and eight different $nd$ states were embedded, and the $|E_b|$ value in which the embedded plasma marker was the same as for the bare plasma was found by interpolation. Secondly, we embedded plasmas with $T_{e,i}$ values in the range 40 to 200 K with $27d$ atoms using a range of $N_R/N_{ion}$ values in different simulations. The specifics of these simulations, and comparisons against the experimental data, are described below.
\subsection{{Modeling the crossover condition}\label{iXMod}} 

We ran simulations of plasmas with $T_{e,i} =$ 20, 30, 40, 50, 60, 80, 100, 120, and 140 K. For each $T_{e,i}$, we performed several sets of runs. First, we did simulations of bare plasmas ($N_R/N_{ion}=0$) with average initial ion densities of $3 \times 10^7$ cm$^{-3}$ and $3 \times 10^8$ cm$^{-3}$. Then, we did simulations of plasmas with $N_R/N_{ion}=0.2$ and 0.3 (for an average initial ion density of $3 \times 10^7$ cm$^{-3}$) and $N_R/N_{ion}=0.3$ (average ion density $3 \times 10^8$ cm$^{-3}$). The simulations used $\sigma_0 = 354 \ \mu$m, which was limited by the maximum computing power given the densities we used. In each of the simulations with $N_R/N_{ion} \ne 0$, a different Rydberg state was embedded (different $|E_b|$ values), and we embedded  between five and eight different $nd$ states for a given $T_{e,i}$, $N_R/N_{ion}$ combination. We identified five critical markers that we used to find the crossover condition: the electron temperature and electron Coulomb coupling parameter at 100 ns of evolution time, $T_{0.1}$ and $\Gamma_{0.1}$ respectively, the electron temperature and coupling parameter at 9.9 $\mu$s, $T_{9.9}$ and $\Gamma_{9.9}$ respectively, as well as $T_{e,0} = m_{ion}v^2_0/k_B$. That is, we find $T_{e,0}$ as defined using Eq. \ref{vZero}, assuming $T_{ion,0} = 0$, and $v_0$ is found from the average rate of change of the characteristic plasma radius $\sigma$ for the period 9.9 to 19.9 $\mu$s of evolution time ($v_0$ changes by a negligible amount over this interval \cite{fore18}). At each $\rho_{ion}$, $N_R/N_{ion}$ combination, the crossover $|E_b|$ value for specific $T_{e,i}$ (20, 30, 40, 50, 60, 80, 100, 120, or 140 K) for a particular marker ($T_{0.1}$, $T_{9.9}$, $\Gamma_{0.1}$, $\Gamma_{9.9}$, $T_{e,0}$) is found using the following method. First, for a given $T_{e,i}$, we fit the values of each of the marker parameters at discrete $|E_b|$ to a cubic spline to obtain values of that marker as a continuous function of $|E_b|$. Then, for each marker, we found the interpolated $|E_b|$ value for which the marker had the same value as the bare plasma with the same $T_{e,i}$ and $\rho_{ion}$. The interpolated $|E_b|$ value is the crossover $|E_b|$ for the $T_{e,i}$ used in the simulation, and we interpret this $T_{e,i}$ to be equivalent to $T_{CO}$ for the interpolated $|E_b|$. Usually, there is no $nd$ state at this $|E_b|$ value, so we assign an estimated uncertainty in $|E_b|$ of one half the adjacent $nd$-state spacing. This corresponds to $\pm 25$ K at $|E_b|/k_B = 500$ K and $\pm 2.5$ K at $|E_b|/k_B = 100$ K. The simulation results for the crossover conditions for the various markers are shown in Fig. \ref{tCO}, along with the experimental results obtained using the electron and ion TOF spectra as described above in Sects. \ref{eTOF} and \ref{iTOFx}.

It should be noted that we find the crossovers by comparing bare plasmas with an average ion density $\rho_{ion}$ with embedded plasmas that have the same average ion density, plus an average Rydberg atom density of $(N_R/N_{ion}) \, \rho_{ion}$. The embedded plasmas therefore have a higher total number of particles than the bare plasmas, though this situation is the same as that in the experiments. We found that the crossover conditions for the markers $T_{0.1}$, $\Gamma_{0.1}$, and $T_{e,0}$ show no significant dependence on $\rho_{ion}$ or $N_R/N_{ion}$, though both of $T_{9.9}$ and $\Gamma_{9.9}$ are quite sensitive to ion density and Rydberg atom to ion ratio. We ran additional simulations in which the crossover $|E_b|$ values for an embedded plasma with $\rho_{ion} = 3 \times 10^8$ cm$^{-3}$ and $N_R/N_{ion} = 0.3$ were found by comparing it with a bare plasma with $\rho_{ion} = 4 \times 10^8$ cm$^{-3}$, and found that the crossover values for $T_{0.1}$, $\Gamma_{0.1}$ and $T_{e,0}$ were the same as those found when the bare plasma ion density was $3 \times 10^8$ cm$^{-3}$. (However, the crossover values for markers $T_{9.9}$ and $\Gamma_{9.9}$ were different when these two situations were compared.)

The experimental results shown in Fig. \ref{tCO} agree with the simulation prediction for the crossovers for the $T_{e,0}$ marker. Specifically, in Sect. \ref{xoverSum}, we summarized the ion and electron results as $|E_b| = 2.7 (\pm 0.5) \times k_B \, T_{CO}$, while the simulation prediction is $|E_b| = 2.9 \times k_B \, T_{CO}$. However, the simulation crossover condition for the $\Gamma_{9.9}$ marker is in the range $2.1 - 2.6 \, \times \, k_B \, T_{CO}$ (this simulation result is sensitive to ion density and $N_R/N_{ion}$). On the other hand, the $T_{0.1}$ and $\Gamma_{0.1}$ crossover simulations substantially agree with each other ($|E_b| \approx 3.5 \times k_B \, T_{CO}$), but are significantly above the experimental results for the $T_{e,0}$ crossovers. Additionally, the simulations for the $T_{9.9}$ crossovers have significant dependence on $\rho_{ion}$ and $N_R/N_{ion}$, but for the conditions that are the most similar to those of the experiment ($\rho_{ion} = 3 \times 10^7$ cm$^{-3}$, $N_R/N_{ion} = 0.3$) the $T_{9.9}$ crossover trend is very similar to the experimental $T_{e,0}$ crossover behavior.

\subsection{{Modeling the amount of heating or cooling}\label{iHeatMod}} 
We also simulated the change in $T_{e,0}$ that occurred when the number of Rydberg atoms added to the UNP was changed. Specifically, we looked at plasmas with average ion densities $\rho_{ion} = 3 \times 10^7$ cm$^{-3}$, $1 \times 10^8$ cm$^{-3}$, and $3 \times 10^8$ cm$^{-3}$, and for each density, we varied $N_R/N_{ion}$ from 0 to 0.3 (for the highest ion density) to as much as 0 to 0.6 (the lowest ion density). The range of $N_R/N_{ion}$ values at a given ion density was again limited by available computing power. At each of the two lower densities, we also looked at the effect of changing $\sigma_0$, though changing this parameter changed the results for a given density only slightly. At each ion density and $N_R/N_{ion}$ value, we evolved plasmas with between four and eight different $T_{e,i}$ values in the range 40 to 200 K. From the results of each simulation, we found $T_{e,0}$ as defined in Eq. \ref{vZero} from the $v_0$ value averaged over the interval 9.9 - 19.9 $\mu$s of plasma evolution time ($v_0$ is found from the change in $\sigma$ over this interval). We then did a linear regression of $T_{e,0}$ versus $T_{e,i}$ to find the slope $\Delta T_{e,0}/\Delta T_{e,i}$, i.e., we used the same protocol as we used for the experimental data as described in Sect. \ref{iTOFx} above. These $\Delta T_{e,0}/\Delta T_{e,i}$ values are plotted versus $N_R/N_{ion}$, along with the experimental data, in Fig. \ref{heating}(b). The vertical error bars of the simulation data in Fig. \ref{heating}(b) are comparable to the size of the line symbols, and are determined solely by the uncertainty in the fit value of the slope of $T_{e,0}$ versus $T_{e,i}$. The simulation results in Fig. \ref{heating}(b) have no horizontal error bars since the $N_R/N_{ion}$ values are input parameters in the simulations. As can be seen, there is good agreement of the experimental data, taken with average ion density $(2.8 \pm 0.7) \times 10^8$ cm$^{-3}$, with the numerical simulations obtained when the ion density is set to $3.0 \times 10^8$ cm$^{-3}$. While we were unable to run the simulations with the same $\sigma_0$ that was used in the experiments (this was $\sigma_0 = 560 \pm 40$ $\mu$m), the variation in the model results as $\sigma_0$ was changed at lower density seen in Fig. \ref{heating}(b) suggests the simulation would be in even stronger agreement with the experimental results had $\sigma_0 = 560$ $\mu$m been used.

\section{{Discussion}\label{disc}} 

There is good agreement of the numerical simulation results with the experiment for both the variation of the crossover temperatures derived from $T_{e,0}$ with $|E_b|$ and for the dependence of $\Delta T_{e,0} / \Delta T_{e,i}$ on $N_R/N_{ion}$ when $27d_{5/2}$ atoms are added to a UNP. In this section, we will consider the underlying plasma processes which lead to these results. 

The interaction of a plasma with a co-existing reservoir of neutral atoms has been considered in many theoretical and numerical studies \cite{mans69,vriens80,steve75,kuz02b,poh08,bann11}. In particular, the programs we have used to obtain the numerical results are derived from the work of Mansbach and Keck (herein abbreviated as MK) \cite{robx02,robx03}. Specifically, the probabilities for Rydberg excitation and de-excitation due to electron collisions used in the Monte-Carlo calculations are based on Eqs. III.12 in Ref. \cite{mans69}, though the programs additionally include the effect of radiative decay, which was considered only briefly by Mansbach and Keck. However, radiative decay plays only a minor role under most of the experimental conditions we used. For all the experimental data points in Fig. \ref{tCO} with $|E_b|/k_B \le 200$ K, the collisional de-excitation rate was at least 10 times larger than the radiative decay rate for the $T_{e,i}$ values and electron densities used, though for the lowest $n$ state we investigated ($n = 24$, for which $|E_b|/k_B = 308$ K), the collisional rate was only 2.4 times the radiative rate \cite{mans69,robx03} at $T_{e,i} = 200$ K. On the other hand, for all the experimental data in Fig. \ref{heating}, the collisional de-excitation rate was at least 30 times the radiative rate (for $n=27$, with average electron density $3 \times 10^8$ cm$^{-3}$, and at the maximum $T_{e,i}$ = 200 K).

\subsection{{Crossover temperature, $T_{CO}$}\label{discXover}} 
If one neglects radiative decay, the equations presented by MK may be used to find the crossover condition where the presence of the embedded Rydberg atoms leaves the plasma unperturbed. Specifically, using Eqs. III.12 in Ref. \cite{mans69}, one can find the mean change in the energy of a Rydberg atom with energy $E_b$ (where $E_b \le 0$) due to electron collisions when the plasma electron temperature is $T_e$. When the collision results in excitation of the Rydberg atom, $\Delta E^e_b = +k_B \, T_e$, but when the collision results in the atom losing energy, $\Delta E^d_b = E_b/2.83 = 0.353 \, E_b$. The $E_b$ value for which there is no net transfer of energy from the Rydberg atoms to the electrons is when $\Delta E_b = \Delta E^d_b + \Delta E^e_b = 0$, which occurs when $E_b = -2.83 \, k_B T_e$. This is very close to what we obtain for the crossover trend in the experimental data and the simulation results when the marker under consideration is $T_{e,0}$. Specifically, the $T_{e,0}$ crossover temperatures obtained from the ion TOF spectra follow the relationship $|E_b| = 2.9 \, k_B T_{CO}$ (when the intercept is unconstrained) and $|E_b| = 2.4 \, k_B T_{CO}$ from the electron TOF spectra.

There are several aspects of this result that warrant further scrutiny. Probably the most significant consideration is the question of why the crossover binding energy of the embedded Rydberg atoms when $T_{e,0}$ is the marker quantity is not the same as the bottleneck energy. A major finding of MK, which has been reproduced in numerous other theoretical analyses \cite{vriens80,steve75,kuz02b,poh08,bann11}, is the existence of a bottleneck in the Rydberg state distribution of atoms in equilibrium with a plasma with electron temperature $T_e$. The energy of the bottleneck given in Ref. \cite{mans69} is $E_{bn} = 3.83 \, k_B T_e$ (other analyses give a slightly different numerical factor), and atoms with binding energy such that $|E_b| < E_{bn}$ are more likely to be excited than de-excited as a result of a single electron-atom collision, and will ultimately ionize, and such collisions will cool the plasma electrons. On the other hand atoms with $|E_b| > E_{bn}$ are likelier to de-excite, and will ultimately decay radiatively or collisionally until their effect on the plasma is negligible. The energy lost by the atom in de-excitation collisions will heat the plasma electrons. One might therefore expect that adding Rydberg atoms with binding energy equal to the bottleneck energy that is characteristic of the electron temperature in the plasma would have no net heating or cooling effect on the plasma, yielding a crossover condition $|E_b| \approx 3.8 \times k_B \, T_{CO}$.

In the simulation results for the crossover condition when the markers $T_{0.1}$ and $\Gamma_{0.1}$ are used, we see a trend ($|E_b| \approx 3.5 \times k_B \, T_{CO}$) that is very similar to that predicted on the basis of the argument that the crossover binding energy is equal to the bottleneck energy. The same result was found in the theoretical analysis presented in Ref. \cite{poh06a}, which considered the effect of adding Rydberg atoms to cold plasmas made from cesium atoms. Specifically, in Fig. 3(b) in Ref. \cite{poh06a}, adding $31d$ ($|E_b|/k_B = 194$ K) atoms to a plasma with $T_{e,i} = 50$ K resulted in no significant change in the plasma electron temperature from that of an unperturbed UNP during the first 600 ns of plasma evolution, although after 600 ns, the electron temperature of the embedded plasma becomes larger than that of the bare plasma. Indeed, by an evolution time of 1 $\mu$s, the electron temperature of the bare plasma is very similar to that of a plasma embedded with $36d$ atoms ($|E_b|/k_B = 141$ K). Adapting these results to our formalism of a crossover condition gives $|E_b| = 3.9 \times k_B \, T_{CO}$ when the marker is the electron temperature from 0 to 600 ns of evolution time ($31d$ result), but $|E_b| = 2.8 \times k_B \, T_{CO}$ when the marker is the electron temperature at 1 $\mu$s of evolution time ($36d$ result). The data shown in Fig. 3(b) in Ref. \cite{poh06a} suggest that the plasma expansion velocity at evolution times $\gg 1 \ \mu$s for a bare plasma would be more similar to that embedded with $36d$ atoms than to a plasma embedded with $31d$ atoms, a finding similar to our own experimental results from the ion TOF spectra, and our simulations when $T_{e,0}$ is the marker quantity.

Our experimental results from the ion TOF spectra, and our simulations, as well as the numerical modeling presented in Ref. \cite{poh06a}, show that the crossover electron temperature for marker $T_{e,0}$ follows the relationship $|E_b| \approx 2.7 \times k_B \, T_{CO}$ rather than what is expected using the argument presented above. However, our result is actually consistent with the bottleneck argument. As described above in Sect.\ref{intro}, there is a distinction between $T_{e,0}$, the effective initial electron temperature which is related to $v_0$ by Eq. \ref{vZero}, and $T_{e,i}$, the electron temperature set by the ionization laser frequency (Eq. \ref{Teinit}). Generally, $T_{e,0}$ is larger that $T_{e,i}$ due to TBR, DIH, and electron-Rydberg scattering. It is this latter process that makes the $T_{e,0}$ crossover condition different from that which is equivalent to adding Rydberg atoms with binding energy equal to the bottleneck energy. Specifically, when one adds Rydberg atoms with $|E_b| = E_{bn} = 3.8 \times k_B \, T_e$, half will ultimately ionize, and half will be scattered into states bound by more than the bottleneck energy. Consequently, there will be an excess population of down-scattered Rydberg atoms with $|E_b| > E_{bn}$. Collisions between the electrons and these atoms result in more down-scattering events than collisions which increase the Rydberg atom energy. Thus, the net effect of the excess atom population with $|E_b| > E_{bn}$ is that the plasma electrons will be heated, and this presumably leads to the increase in the electron temperature after 600 ns seen in Fig. 3(b) in Ref. \cite{poh06a} and to an increase in $T_{e,0}$. On the other hand, when one adds Rydberg atoms with $|E_b| = 2.8 \times k_B \, T_e$, more than half will ionize, but the average energy of the down-scattered Rydberg atoms is $E_b + \Delta E_b = E_b + 0.353 E_b = -3.8 \times k_B \, T_e$, i.e., the same as the bottleneck energy for a plasma with electron temperature $T_e$. Subsequent scattering events of electrons with the resulting Rydberg reservoir with average energy $- E_{bn}$ will change the plasma electron temperature only if the population distribution of the down-scattered embedded Rydberg atoms is significantly different from the distribution in a bare plasma. 

A second consideration with regard to our experimental results, and our simulations, is why there is a difference in the bottleneck condition when $T^\ast_{e,0}$ derived from the electron TOF spectra is used as the marker, rather than $T_{e,0}$ from the ion TOF spectra. This is likely due to the fact that the model described in Ref. \cite{twed12} which we used to obtain the $v^\ast_0$ proxy as described in Sect. \ref{eTOF}, assumes the electrons are a zero-temperature fluid, i.e., they just leak across the top of the barrier formed by the average potential due to the ions and electrons in the UNP and the external electric field. This is clearly unrealistic: The electrons evaporate out of the plasma, and so the $v^\ast_0$ proxy is likely more similar to $\Gamma_e$ in its behavior than it is to $v_0$, since $\Gamma_e$ is the ratio of the mean electron electrostatic interaction energy to the electron thermal kinetic energy. Indeed, the crossover behavior obtained from the electron TOF spectra has some similarity to the simulation results for the marker $\Gamma_{9.9}$, though the crossover values found using this quantity (like those using $T_{9.9}$) are very sensitive to the ion density and $N_R/N_{ion}$ values used in the simulations.

To summarize, we find the results of the experiments and the numerical simulations of the crossover condition for the marker $T_{e,0}$ are well explained using the equations for excitation and de-excitation of Rydberg atoms due to electron collisions presented in Ref. \cite{mans69}. Specifically, in the experiments the crossover between cooling and heating of a plasma with an initial electron temperature $T_{e,i}$ by net energy transfers to or from the Rydberg reservoir is when the binding energy is such that $|E_b| \approx 2.7 \times k_B T_{e,i}$. The corresponding result from the numerical simulations is $|E_b| \approx 2.9 \times k_B T_{e,i}$. Using the equations in MK \cite{mans69}, the condition for zero net exchange of energy between the Rydberg atoms and the plasma electrons with temperature $T_e$ is $|E_b| = 2.83 \times k_B T_{e}$. This latter condition does not change significantly when revised electron-atom collision rate equations from a more recent study by Pohl \textit{et al.}, are used, as described below in Sect. \ref{discHeat} \cite{poh08}.

\subsection{{Amount of heating}\label{discHeat}} 

We can also use the mean energy transfer to or from the plasma electrons caused by electron-Rydberg collisions from MK to understand the experimental results described in Sect. \ref{ieTOFheat}. Specifically, the mean amount of heating of the plasma electrons that results from a single de-excitation collision is $0.353 \, |E_b|$, and the mean amount of cooling due to a single atom excitation collision is $k_B \, T_e$. Thus, the net thermal energy transferred to the plasma by a single electron-Rydberg collision is
\begin{equation}
\Delta E_{th} = 0.353 \, |E_b| - k_B \, T_e = k_B \, (T_{CO} - T_e), \label{netHeat}
\end{equation}
where we have used the equality $k_B \, T_{CO} = 0.353 \, |E_b|$. That is, $T_{CO}$ is the crossover temperature appropriate to the particular Rydberg state being embedded, and we are assuming that the appropriate crossover marker is that for $T_{e,0}$ since we are looking at the effect that adding Rydberg atoms to the UNP has on the plasma expansion velocity. 

We can now estimate the effect of these collisions on the plasma expansion. We assume an initial situation where we have $N_{ion,i}$ electrons in the plasma at an initial temperature $T_{e,i}$, giving an initial electron thermal energy of $E_i = N_{ion,i} \, \frac{3}{2} \, k_B \, T_{e,i}$. Our argument neglects the thermal energy of the ions, since the maximum ion temperature is of order 1 K \cite{kill07}. We also do not explicitly include the binding energy of the Rydberg atoms; rather, we just consider the extent to which the change in the energy of the Rydberg atoms heats the electrons via Eq. \ref{netHeat}. The electrons are heated by collisions with an initial number Rydberg atoms $N_{R}$ over the course of the plasma evolution, resulting in a final situation late in the plasma evolution when almost all of the electron thermal energy has been converted to radial outward motion of the ions. We will assume that the number of Rydberg atoms which ionize is $\Delta N_R$, so that the number of electrons at the end of the evolution is $N_{ion,f} = N_{ion,i} + \Delta N_R$. The final energy of the system in this case is
\begin{eqnarray}
E_f &=& E_i \, + \, \eta_{coll} N_R \Delta E_{th} \nonumber \\
&=& N_{ion,f} \frac{3}{2} k_B T_{e,f} + N_{ion,f} \frac{3}{2} m_{ion} \gamma^2 \sigma^2, \label{eFinal}
\end{eqnarray}
where the $\eta_{coll}$ is the number of electron collisions each Rydberg atom experiences during the plasma evolution and the last term on the second line is the kinetic energy due to the radial motion of the ions late in the plasma evolution. (The parameter $\gamma$ relates an ion's outward velocity at time $t$ to its position relative to the center of the plasma, $\vec r$, $\vec u(\vec r,t) = \gamma \, \vec r$ \cite{kill07,poh04b}.) Averaged over the Gaussian spatial distribution of the ions, and in the limit where $t \rightarrow \infty$, $m_{ion} \, \gamma^2 \, \sigma^2 \rightarrow m_{ion} \, v^2_0 = k_B \, T_{e,0}$ \cite{kill07,fore18}. 

Strictly, Eq. \ref{eFinal} is valid only in the limit where $\eta_{coll} \ll 1$, which would leave the Rydberg state distribution relatively unaffected by collisions with the plasma electrons. Additionally, if $\eta_{coll} \ll 1$, the electron temperature change is small during the period in which the electron-Rydberg atom collisions occur, so one can change $T_e$ from Eq. \ref{netHeat} to be $T_{e,i}$ in the $\Delta E_{th}$ term in Eq. \ref{eFinal}. (The period in which most of the electron-Rydberg atom collisions occur is limited to the first few $\mu$s of plasma evolution - as the plasma expands, the electron density in the central region falls rapidly, causing the collision rate to fall.) We will consider the validity of the assumption $\eta_{coll} \ll 1$, and the consequences when it is not valid, below. Finally, we will neglect the $k_B \, T_{e,f}$ term in Eq. \ref{eFinal} in comparison with the $k_B \, T_{e,0}$ term. That is, we are assuming adiabatic expansion has converted almost all of the electron thermal energy into outward motion of the ions. 

Solving Eq. \ref{eFinal} with these assumptions, and using Eq. \ref{netHeat}, we obtain the relationship
\begin{equation}
T_{e,0} = \frac{N_{ion,i}}{N_{ion,f}} \, T_{e,i} + \frac{2}{3} \, \eta_{coll} \, \frac{N_R}{N_{ion,f}} \, (T_{CO} - T_{e,i}). \label{bigeq}
\end{equation} 
We will make one further simplification, which is to assume that a negligible fraction of the embedded Rydberg atoms ionize, so we can set $N_{ion,i} = N_{ion,f} = N_{ion}$. (After making this approximation, it is immediately obvious that Eq. \ref{bigeq} reproduces the experimental behavior described by Eq. \ref{expAmtHeat}.) Obviously, this approximation is very good when $N_R = 0$, since the difference between $N_{ion,i}$ and $N_{ion,f}$ in this case is due only to TBR. Variations in the ratios ${N_{ion,i}}/{N_{ion,f}}$ and ${\eta_{coll}}/{N_{ion,f}}$ with $T_{e,i}$ when $N_R \ne 0$ certainly occur in the numerical simulations, and the error introduced by assuming that the ion number is constant is on the order of 20\% at the highest ion density and $N_R/N_{ion}$ ratio we looked at. However, we cannot decouple changes in ${N_{ion,i}}/{N_{ion,f}}$ and ${\eta_{coll}}/{N_{ion,f}}$ when $N_R$ is changed in the experimental results. It therefore makes sense to assume $N_{ion}$ is constant, since doing so does it alter the major conclusions of our analysis. 

Expressing Eq. \ref{bigeq} as a derivative relationship assuming constant $T_{CO}$ and no significant variation of $\eta_{coll}$ with $T_{e,i}$, we obtain
\begin{equation}
\frac{\Delta T_{e,0}}{\Delta T_{e,i}} = 1 - \frac{2}{3} \, \eta_{coll} \, \frac{N_R}{N_{ion}}. \label{slopeEq}
\end{equation}
For bare plasmas, $N_R = 0$, and thus $N_{ion,f} = N_{ion,i}$ (there will be a small difference in the two due to TBR which we ignore), so ${\Delta T_{e,0}}/{\Delta T_{e,i}} = 1$. For embedded plasmas $N_R > 0$, but any effect due to $N_{ion,f}$ being different from $N_{ion,i}$ is masked because the second term on the right side of Eq. \ref{slopeEq} is non-zero. Additionally, the term ``$\eta_{coll}$'' in Eq. \ref{slopeEq} should be $\eta_{coll} \, N_{ion}/N_{ion,f}$, but again, variations in this latter term that depend on $N_R$ or $T_{e,i}$ cannot be isolated from the experimental results.

Before we compare Eq. \ref{slopeEq} with the results shown in Fig. \ref{heating}, it is worth considering two assumptions about $\eta_{coll}$ that were used in deriving the equation. The first is that $\eta_{coll}$ has no significant dependence on $T_{e,i}$, and the second is that $\eta_{coll} \ll 1$. Mansbach and Keck give an equation for the total collision rate of electrons with each Rydberg atom in an initial state with energy $E_i$ when the electron temperature is $T_e$, $k(\epsilon_i)$, where $\epsilon_i = E_i/k_B T_e$  (Eq. III.14 in Ref. \cite{mans69}). For our experiment we use $T_e = T_{e,i}$ values $40 \ \textrm{K} \le T_{e,i} \le 200$ K and $E_i = E_b$ such that $|E_b|/k_B = 240$ K (for the embedded $27d$ atoms), and for this range of parameters, $k(\epsilon_i)$ increases by a factor of 2 as $T_{e,i}$ is changed from 40 K to 200 K at fixed electron density. On the other hand, as $T_{e,i}$ increases, the electron density falls more quickly. Specifically, at the center of the UNP, the time for the density to fall to 50\% of the initial value is $\sqrt{(2^{2/3}-1)} \, \sigma_0/v_0 \propto 1/\sqrt{T_{e,i}}$ (strictly, the proportionality is valid only if there is no heating or cooling of the electrons). The number of electron collisions experienced by each Rydberg atom during the time the density falls to half its initial value is thus constant within 20\% for $T_{e,i} = 40 \ \textrm{K} \le T_{e,i} \le 200$ K, and each atom experiences $\approx 3$ collisions in this time for an initial electron density of $3 \times 10^8$ cm$^{-3}$. Our assumption that $\eta_{coll} \ll 1$ is not valid for the experimental data shown in Fig. \ref{heating}, though it likely is acceptable for the lowest density we ran in the simulations shown in the figure, $3 \times 10^7$ cm$^{-3}$. 

There are two main consequences when $\eta_{coll} \gtrsim 1$ with regard to Eq. \ref{slopeEq}. First, the binding energy for most atoms will differ significantly from the initial value, and so it cannot be assumed that $T_{CO}$ in Eq. \ref{bigeq} is constant. Second, the electron temperature will change significantly from $T_{e,i}$. The term we have written as $T_{CO} - T_{e,i}$ in Eq. \ref{bigeq} is actually some sort of ensemble average of $0.353 \, |E_{b}(t)|/k_B - T_e(t)$, where $E_b(t)$ is the Rydberg binding energy at time $t$, and $T_e(t)$ is the corresponding electron temperature. However, the net effect of electron-Rydberg atom collisions is that $0.353 \, |E_{b}(t)|/k_B - T_e(t) \rightarrow 0$, as can be seen in the data in Figs. \ref{iAndeTOFspectra} and \ref{heating}(a). Practically, this means that the second term on the right hand side of Eq. \ref{slopeEq} cannot exceed unity in magnitude (i.e., the maximum amount of heating or cooling is limited by the $T_{CO}$ value set by the binding energy of the added Rydberg atoms), and when comparing Eq. \ref{slopeEq} with the experimental data and simulations, the value of $\eta_{coll}$ obtained will be a significant underestimate in the region where ${\Delta T_{e,0}}/{\Delta T_{e,i}} \rightarrow 0$.

Despite these significant simplifications, Eq. \ref{slopeEq} describes well the behavior we see in the experimental data and the numerical simulations for the dependence on the slope of the $T_{e,0}$ versus $T_{e,i}$ graph with $N_R/N_{ion}$. The experimental data points, obtained at an average ion density of $\rho_{ion} = (2.8 \pm 0.7) \times 10^8$ cm$^{-3}$ agree with the numerical simulations for a density of $3.0 \times 10^8$ cm$^{-3}$ within the experimental uncertainty, and when comparing simulation results at lower density for different $\sigma_0$, the agreement would be even better had the simulations been run at the same $\sigma_0$ value as for the experiment. For the experimental data shown in Fig. \ref{heating}(b), using Eq. \ref{slopeEq}, the slope of $\Delta T_{e,0}/\Delta T_{e,i}$ with $N_R/N_{ion}$ is $-1.6 \pm 0.3$, which suggests a value of $\eta_{coll} = 2.4 \pm 0.5$. However, as noted above, this is an underestimate due to the fact that the mean Rydberg atom energy and the electron temperature change during the co-evolution of the UNP and the Rydberg reservoir. Again, this is in line with the estimate of $\eta_{coll} \approx 3$ during the time the electron density falls by 50\% based on MK Eq. III.14. In comparing the simulation results obtained using $\sigma_0 = 354$ $\mu$m and the three different initial ion densities in the ratio 1 : 3.3 : 10 (the initial ion density is the same as the initial electron density), the slopes of the graphs are in the ratio 1 : 3.0 : 7.2, showing that successive electron-Rydberg atom collisions become less effective at heating or cooling the plasma when the density increases due to the convergence of the mean Rydberg atom energy and the electron temperature. Finally, the slope of the simulation result for an average ion density of $3 \times 10^7$ cm$^{-3}$ (for this density, we expect Eq. \ref{slopeEq} will have the greatest accuracy of all the densities we simulated) and $\sigma_0 = 566 \ \mu$m gives $\eta_{coll} \approx 0.4$, which is similar to the value we get using MK Eq. III.14 at this density in the time density drops to half its initial value (maximum and minimum values $\eta_{coll} = 0.34$ at $T_{e,i} = 40$ K, $\eta_{coll} = 0.29$ at $T_{e,i} = 130$ K).

\subsection{{Sensitivity of numerical results to model assumptions}\label{discHeat}} 

The work of Mansbach and Keck is now almost 50 years old. A more recent numerical study by Pohl \textit{et al.} \cite{poh08} found that the MK equations significantly underestimate the rate of small energy transfer collisions, although they also found the MK rates to be accurate for moderate to large energy transfer collisions. We have tested the robustness of our findings regarding the crossover temperature and the amount of heating or cooling when Rydberg atoms are added to a UNP by estimating the mean change in the Rydberg atom energy based on the equations of Pohl \textit{et al.} The strong maxima in the excitation and de-excitation collision rates in the region $|\Delta E_R|/k_B T_e \lesssim 0.1$ (where $\Delta E_R$ is the change in the Rydberg atom energy) reduce significantly both the average increase in Rydberg atom energy (and consequent cooling effect on the UNP) and average decrease in Rydberg atom energy due to collisions. For example, for a Rydberg atom with same binding energy as the $27d$ state ($|E_b|/k_B = 240$ K) and $T_e = 100$ K, the mean energy increase is $\approx 0.13 \, k_B T_e$ (rather than the MK prediction of $k_B T_e$), and the mean decrease is $\approx 0.042 \, |E_b|$ (the MK prediction is $0.35 \, |E_b|$). However, the up/down energy changes are equal at a crossover temperature such that $|E_b| = 3.2 \times k_B T_{CO}$, rather than the MK prediction of $|E_b| = 2.8 \times k_B T_{CO}$. The crossover prediction based on the rates in Ref. \cite{poh08} change with $T_e$, and for $T_e = 40$ K, one obtains $|E_b| = 2.0 \times k_B T_{CO}$. (On the other hand, the multiplication factor changes only from 2.9 to 3.3 over the range $80 \le T_e \le 200$ K.) The primary consequences with regard to our study of the collision rates reported by Pohl \textit{et al.} being different from those given by MK are that $T_{CO}$ should depend in some way on $N_R/N_{ion}$, and that for a given $N_R/N_{ion}$, $T_{CO}$ should be found in the experiment by incrementing $T_{e,i}$ until the condition $\Delta T_{e,0} = 0$ is found, rather than doing a linear regression of $\Delta T_{e,0}$ with $T_{e,i}$. However, given the experimental uncertainty inherent in the measurement techniques used to obtain the data in Figs. \ref{tCO} and \ref{heating}, our experimental $T_{CO}$ values and associated uncertainty ranges likely encompass any of the potential variations that are manifestations of effects based on the collision rate equation in Ref. \cite{poh08}. In other words, our experimental results are consistent with both the MK collision rate equations as well as those of Pohl \textit{et al.} Using the equations of Pohl \textit{et al.}, more collisions are needed to give a certain amount of heating or cooling, but they also predict higher collision rates for small energy transfers than MK anyway, so the fundamental prediction is the same as that based on the equations presented by MK \cite{mans69}.

\section{CONCLUSION}

We have demonstrated experimentally both heating and cooling of electrons in UNPs using Rydberg atoms which are embedded in the plasma at its creation. The experimental behavior of the crossover between heating and cooling, and the amount of heating or cooling that can be achieved, are in agreement with the results of Monte-Carlo simulations. In particular, a significant degree of cooling can be achieved. In our experiment, when we added $27d$ atoms to plasmas with $T_{e,i} = 200$ K at an initial ion density of $3 \times 10^8$ cm$^{-3}$, the effective initial electron temperature changed to $T_{e,0} \approx 150$ K, a decrease of $\approx$30\% from the value for a bare plasma. While the simulations for this situation showed a more modest decrease in $T_{e,0}$ of $\approx$20\%, the numerical modeling results show that the changes in $T_{e,0}$ are correlated with changes in $\Gamma_e$. For instance, in the experiments where we added $27d$ atoms to UNPs with $T_{e,i} = 200$ K when $N_R/N_{ion} = 0.3$ at an ion density of $3 \times 10^8$ cm$^{-3}$, the Coulomb coupling parameter at an evolution time of $t = 100$ ns, $\Gamma_{0.1}$, increased from 0.009 to 0.0096, and at $t = 9.9 \ \mu$s, $\Gamma_{9.9}$ increased from 0.037 to 0.044. (For the same conditions with $T_{e,i} = 40$ K, the $\Gamma_e$ values decreased by similar percentages.) 

Unfortunately, while the decrease on $T_{e,0}$ values we have observed experimentally and the corresponding increase in $\Gamma_e$ values seen in the simulations when Rydberg atoms are added to cold plasmas are significant, the utility of the technique in reaching the strongly coupled regime for electrons in a UNP is not demonstrated by our results. The most favorable situation we investigated in the simulations was adding $95d$ atoms ($|E_b|/k_B = 18$ K) to UNPs with $T_{e,i} = 20$ K, for which $\Gamma_{0.1}$ increased from 0.09 to 0.15 when $N_R/N_{ion} = 0.3$ and $\rho_{ion} = 3 \times 10^8$ cm$^{-3}$. To date, one of the largest experimental $\Gamma_e$ values that has been reported is $\Gamma_e = 0.35$ for a UNP with $T_{e,i} = 0.1$ K and $\rho_{ion} = 6 \times 10^6$ cm$^{-3}$ \cite{chen17}. It is possible that our technique could reach a comparable $\Gamma_e$ starting from a higher $T_{e,i}$ than 0.1 K and using a higher density. However, this would increase TBR, and it is not clear that adding Rydberg atoms to a UNP could significantly counteract this source of heat for plasma electrons. Additionally, the smaller the values of $T_{e,i}$ would require smaller $|E_b|$ values for the added Rydberg atoms, given the requirement that $|E_b| \lesssim 2.7 \times k_B \, T_e$ for the atoms to achieve cooling of the plasma electrons. Finally, as the value of $|E_b|$ gets smaller, such atoms would spontaneously evolve to plasma more rapidly, even at low density. This would inevitably lead to plasmas in which the electron properties are much more significantly determined by the parent Rydberg atom ensemble, rather than than those of the UNP created by photoionization. In particular, such Rydberg plasmas seem to have a fundamental limit where $\Gamma_e \lesssim 0.1$ \cite{fore18} for densities in the range $10^7$ to $10^9$ cm$^{-3}$.

\section{ACKNOWLEDGEMENTS}

The following former Colby undergraduates worked on preliminary experiments to those described here: J. L. Carini, L. P. Rand, C. Vesa, and R. O. Wilson. In addition, significant improvements to the apparatus were made by D. B. Branden and S. E. Galica. 
We are deeply indebted to K. A. Twedt for sharing the program described in Ref. \cite{twed10} and advice on how to run it. In addition, we acknowledge extensive discussions with T. F. Gallagher, who first suggested this experiment and provided an equipment loan, C. W. S. Conover, and A. L. Goodsell. Preliminary experiments were carried out with an equipment loan from D. M. Thamattoor. D. A. T. acknowledges support from Colby College through the Division of Natural Sciences grants program, and from the National Science Foundation (1068191). F. Robicheaux was supported in part by the US
Department of Energy (DE-SC0012193).

\end{document}